\documentclass[twocolumn]{aastex631}

\usepackage{graphicx}
\usepackage{amsmath}	
\usepackage{amssymb}
\usepackage{booktabs}
\usepackage{enumerate}
\usepackage{dirtytalk}
\usepackage{float}
\usepackage{enumitem}
\usepackage{calligra}
\usepackage{booktabs}
\usepackage{lipsum} 
\usepackage{float}
\usepackage{xcolor}
\usepackage{multirow}
\usepackage{makecell}
\usepackage{blindtext}
\usepackage{subfigure}

\shorttitle{Quartz Clouds on HD189733 b}
\shortauthors{Inglis et al.}
\graphicspath{{./}{figures/}}
\begin{document}

\title{Quartz Clouds in the Dayside Atmosphere of the Quintessential Hot Jupiter HD~189733~b}

\author[0000-0001-9164-7966]{Julie Inglis}
\affiliation{Division of Geological and Planetary Sciences, California Institute of Technology, Pasadena, CA, 91125, USA}

\author[0000-0003-1240-6844]{Natasha E. Batalha}
\affiliation{NASA Ames Research Center, Moffett Field, CA 94035, USA}

\author[0000-0002-8507-1304]{Nikole K. Lewis}
\affiliation{Department of Astronomy and Carl Sagan Institute, Cornell University, 122 Sciences Drive, Ithaca, NY 14853, USA}

\author[0000-0003-3759-9080]{Tiffany Kataria}
\affiliation{Jet Propulsion Laboratory, California Institute of Technology, 4800 Oak Grove Drive, Pasadena, CA 91109, USA}

\author[0000-0002-5375-4725]{Heather A. Knutson}
\affiliation{Division of Geological and Planetary Sciences, California Institute of Technology, Pasadena, CA, 91125, USA}

\author[0000-0003-4220-600X]{Brian M. Kilpatrick}
\affiliation{National Nuclear Security Administration, U.S. Department of Energy, Washington D.C., 20585, USA}

\author[0009-0003-2576-9422]{Anna Gagnebin}
\affiliation{Department of Astronomy and Astrophysics, University of California, Santa Cruz, CA 95064, USA}

\author[0000-0003-1622-1302]{Sagnick Mukherjee}
\affiliation{Department of Astronomy and Astrophysics, University of California, Santa Cruz, CA 95064, USA}

\author[0000-0002-4732-960X]{Maria M. Pettyjohn}
\affil{School of Chemistry, University of New South Wales, 2052, Sydney, Australia}

\author{Ian J.\ M.\ Crossfield}
\affiliation{Department of Physics and Astronomy, University of Kansas, Lawrence, KS, USA}

\author[0000-0002-6276-1361]{Trevor O. Foote}
\affiliation{Department of Astronomy and Carl Sagan Institute, Cornell University, 122 Sciences Drive, Ithaca, NY 14853, USA}

\author[0000-0001-5878-618X]{David Grant}
\affiliation{HH Wills Physics Laboratory, University of Bristol, Tyndall Avenue, Bristol, BS8 1TL, UK}

\author[0000-0003-4155-8513]{Gregory W. Henry}
\affiliation{Tennessee State University (retired), Nashville, TN 37216}

\author[0000-0002-4443-6725]{Maura Lally}
\affiliation{Department of Astronomy and Carl Sagan Institute, Cornell University, 122 Sciences Drive, Ithaca, NY 14853, US}

\author[0000-0003-1039-2143]{Laura K. McKemmish}
\affil{School of Chemistry, University of New South Wales, 2052, Sydney, Australia}

\author[0000-0001-6050-7645]{David K. Sing}
\affil{Department of Earth \& Planetary Sciences, Johns Hopkins University, Baltimore, MD, USA}
\affil{Department of Physics \& Astronomy, Johns Hopkins University, Baltimore, MD, USA}

\author[0000-0003-4328-3867]{Hannah R. Wakeford}
\affiliation{HH Wills Physics Laboratory, University of Bristol, Tyndall Avenue, Bristol, BS8 1TL, UK}

\author[0000-0003-3916-9850]{Juan C. Zapata Trujillo}
\affil{School of Chemistry, University of New South Wales, 2052, Sydney, Australia}

\author[0000-0012-3245-1234]{Robert T. Zellem}
\affiliation{Jet Propulsion Laboratory, California Institute of Technology, 4800 Oak Grove Drive, Pasadena, CA 91109, USA}

\begin{abstract}
Recent mid-infrared observations with JWST/MIRI have resulted in the first direct detections of absorption features from silicate clouds in the transmission spectra of two transiting exoplanets, WASP-17 b and WASP-107 b. In this paper, we measure the mid-infrared ($5-12$~$\mu$m) dayside emission spectrum of the benchmark hot Jupiter HD~189733~b with MIRI LRS by combining data from two secondary eclipse observations. We confirm the previous detection of H$_2$O absorption at 6.5~$\mu$m from Spitzer/IRS and additionally detect H$_2$S as well as an absorption feature at 8.7~$\mu$m in both secondary eclipse observations. The excess absorption at 8.7~$\mu$m can be explained by the presence of small ($\sim$0.01~$\mu$m) grains of SiO$_2$[s] in the uppermost layers of HD~189733~b's dayside atmosphere. This is the first direct detection of silicate clouds in HD~189733~b's atmosphere, and the first detection of a distinct absorption feature from silicate clouds on the day side of any hot Jupiter. We find that models including SiO$_2$[s] are preferred by $6-7\sigma$ over clear models and those with other potential cloud species. The high altitude location of these silicate particles is best explained by formation in the hottest regions of HD~189733~b's dayside atmosphere near the substellar point.
We additionally find that HD~189733~b’s emission spectrum longward of 9 $\mu$m displays residual features not well captured by our current atmospheric models.
When combined with other JWST observations of HD~189733~b's transmission and emission spectrum at shorter wavelengths, these observations will provide us with the most detailed picture to date of the atmospheric composition and cloud properties of this benchmark hot Jupiter. 
\end{abstract}


\section{Introduction} \label{sec:intro}

HD~189733~b is unique amongst the population of hot Jupiter planets detected in the last few decades. At a distance of 19 pc, it is one of the closest of these planets, orbiting a bright (V=7.7) K star with an orbital period of only 2.219 days \citep{bouchy_elodie_2005}. This makes HD~189733~b one of the most favorable hot Jupiter targets for detailed characterization studies. It was also one of the earliest transiting hot Jupiters discovered, and as a result it has accumulated a long list of historic firsts in the field of exoplanet atmospheric characterization. 

Spitzer observations of HD~189733~b provided one of the earliest direct measurements of thermal emission from a transiting exoplanet, with a detection of its 16~$\mu$m secondary eclipse by \cite{deming_strong_2006} using the Infrared Spectrograph (IRS). This planet was also one of the only hot Jupiters to have its mid-infrared emission spectrum measured with IRS \citep{grillmair_spitzer_2007,grillmair_strong_2008,todorov_updated_2014}. HD~189733~b was the first planet to have its surface brightness mapped using a partial 8~$\mu$m Spitzer phase curve \citep{knutson_map_2007}, and the first to have a complementary dayside brightness map constructed from an ensemble of 8~$\mu$m secondary eclipses \citep{majeau_two-dimensional_2012,de_wit_towards_2012}. Both the phase curve measurements and eclipse map revealed that the hottest region in HD~189733~b's atmosphere is offset from the substellar point, indicating the presence of strong eastward jets that also warm the planet's permanently shadowed night side. HD~189733~b has been extensively studied both in transmission \citep[e.g.,][]{pont_hubble_2007,sing_transit_2009,morello_new_2014,mccullough_water_2014} and emission \citep[e.g.,][]{deming_strong_2006,knutson_map_2007,charbonneau_broadband_2008,agol_climate_2010,crouzet_water_2014}, from both spaced-based observatories and ground-based telescopes. This extensive body of observations has made HD~189733~b into a benchmark system for studies of other hot Jupiters.

\begin{figure*}[t!]
\epsscale{1.1}
\plotone{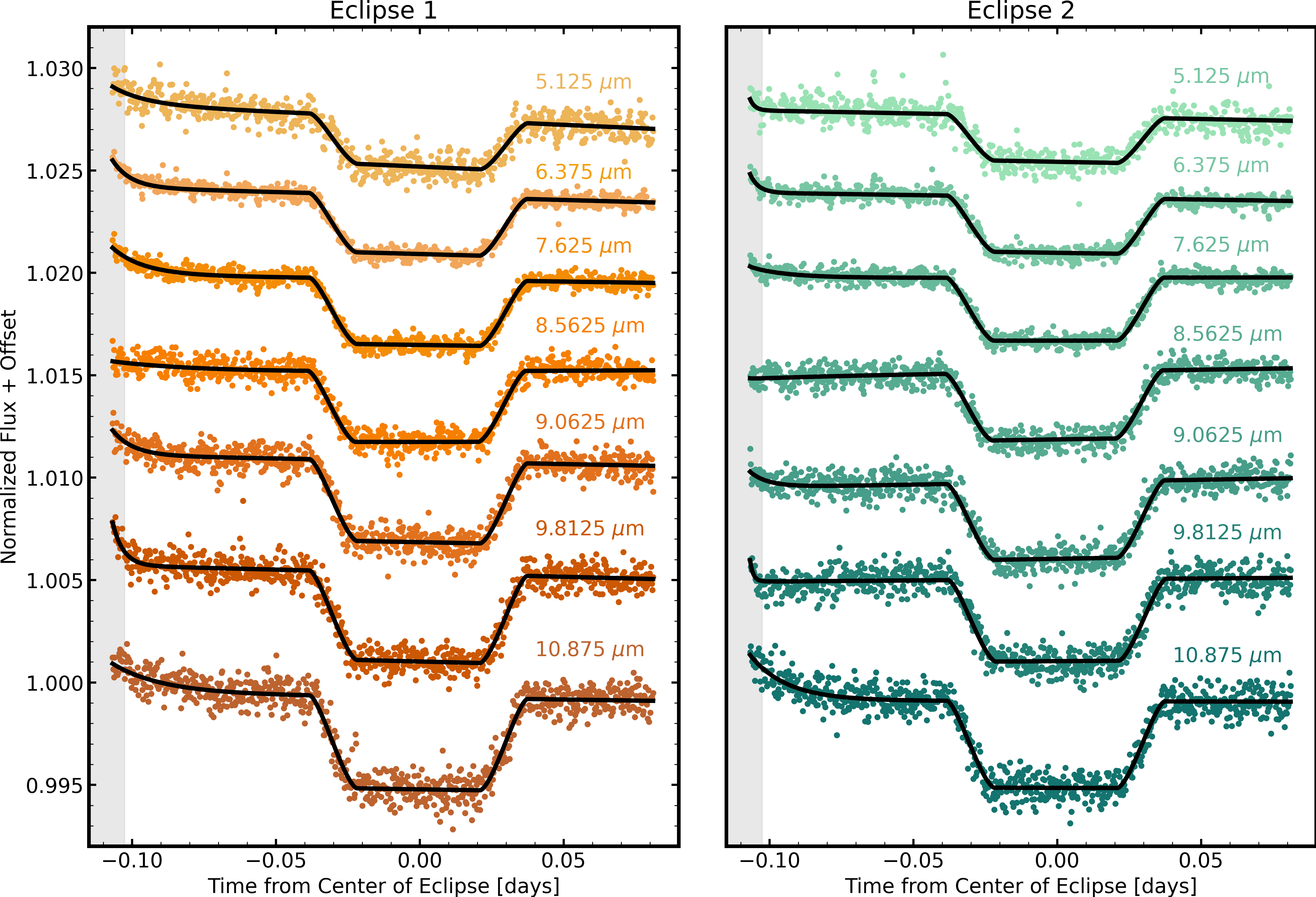}
\caption{A sample of spectroscopic light curves from our \texttt{Eureka!} reduction for each of our two secondary eclipse observations. For our fits, we trim 100 integrations at the beginning of each observation to remove the sharp ``hook'' which cannot quite be captured by an exponential, shown in the grey shaded region. The best-fit combined instrumental and eclipse models are overplotted in black. 
\label{fig:LC_correct}}
\end{figure*} 

HD~189733~b has been the subject of many atmospheric retrieval studies over the years, both in transmission and emission \citep[e.g.,][]{madhusudhan_h2o_2014,welbanks_massmetallicity_2019,zhang_platon_2020,finnerty_atmospheric_2023}. These retrievals generally relied on a combination of spectroscopic detections of water absorption at near-infrared wavelengths and photometric constraints on the abundances of methane, carbon monoxide, and carbon dioxide at longer wavelengths. Very early on, water absorption was detected in HD~189733~b's mid-infrared dayside emission spectrum using Spitzer IRS observations \citep{charbonneau_broadband_2008,grillmair_strong_2008,todorov_updated_2014}. Although early transmission spectroscopy with HST/NICMOS found a near featureless spectrum at near-infrared wavelengths \citep{sing_transit_2009}, more precise observations with HST/WFC3 later revealed a water absorption feature at 1.4~$\mu$m \citep{mccullough_water_2014}. A comprehensive joint retrieval on HD~189733~b's HST and Spitzer transmission and emission spectra found a modestly enhanced atmospheric metallicity of $12^{+8}_{-5}$ times solar and an atmospheric C/O ratio of $0.66^{+0.05}_{-0.09}$ \citep{zhang_platon_2020}. Most recently, \cite{finnerty_atmospheric_2023} performed retrievals on high-resolution ground-based emission observations and found a lower C/O ratio of $0.3\pm0.1$ (2.7$\sigma$ difference) and a consistent atmospheric metallicity [C/H] of $1-8\times$ solar. They also measure a log water abundance of $-2.0\pm0.4$ (mass-mixing ratio). HD~189733~b has also been observed in both transmission and emission by four separate programs during JWST's Cycle 1; these observations will significantly improve our knowledge of its atmospheric composition.

Early transmission spectroscopy of HD~189733~b with HST also provided some of the earliest evidence for the presence of clouds or hazes in a planetary atmosphere. This planet's transmission spectrum exhibits a large optical scattering slope and muted near-infrared water absorption feature, indicating the presence of small scattering particles in the day-night terminator region \citep{PONT_2008,lecavelier_des_etangs_rayleigh_2008,sing_transit_2009,sing_hubble_2011,madhusudhan_h2o_2014}. \citet{wakeford_transmission_2015} identify scattering from small, sub-micron grains of enstatite (MgSiO$_3$[s]) as a potential source. Although HD~189733 is a relatively active K star, the super-Rayleigh scattering slope observed at optical wavelengths cannot be explained by the presence of unocculted star spots alone and is best reproduced by high altitude sub-$\mu$m scattering particles \citep{sing_hubble_2011}.

The presence of clouds in HD~189733~b's dayside atmosphere has also been more controversially inferred through measurements of its optical albedo. \cite{evans_deep_2013} published a geometric albedo measurement using HST STIS that showed a sharp increase in reflected light at bluer wavelengths, suggesting the presence of sub-$\mu$m scattering aerosols. Although measurements of HD~189733~b's albedo from ground-based polarimetry initially found evidence for a significantly higher albedo \citep{berdyugina_polarized_2011}, subsequent observations by \cite{wiktorowicz_ground-based_2015} resulted in a limit consistent with the lower albedo value from \cite{evans_deep_2013}. More recently, CHEOPS measured a smaller geometric albedo than found by \cite{evans_deep_2013} using optical broadband photometry, consistent with predictions from dayside atmosphere models without significant scattering from clouds \citep{krenn_geometric_2023}. A $\sim3\times$ solar Na abundance is required to explain both the CHEOPS and \cite{evans_deep_2013} observations.

The current ensemble of evidence therefore suggests that HD~189733~b likely hosts aerosols in its day-night terminator region, and these aerosols may also be abundant enough in the uppermost layers of the dayside atmosphere to significantly alter its optical albedo. Although we do not currently have any direct constraints on the composition of these aerosols, we can nonetheless make an educated guess based on HD~189733~b's measured atmospheric temperature distribution. Silicate clouds are expected to be the dominant source of cloud opacity at the temperatures and pressures relevant for most hot Jupiter atmospheres and to settle out of the visible atmosphere at temperatures just below the expected dayside temperature of HD~189733~b \citep{gao_aerosol_2020}. However, general circulation models incorporating cloud formation and transport processes indicate that these particles can be carried far from their initial formation locations \citep{woitke_dust_2003,parmentier_3d_2013}.

\begin{figure*}[t!]
\epsscale{1.2}
\plotone{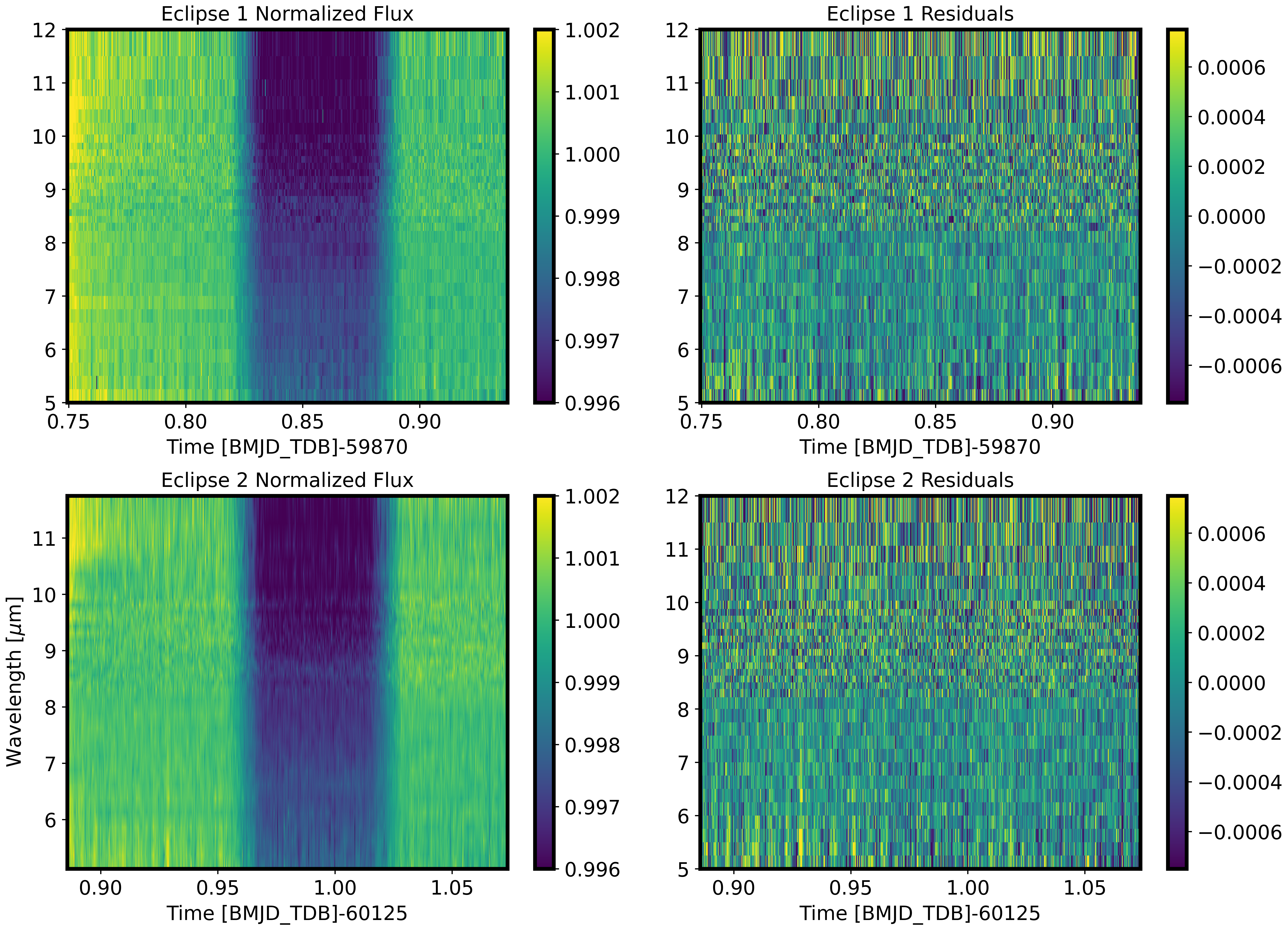}
\caption{(left) Raw spectroscopic light curves from the \texttt{Eureka!} reduction for each of our MIRI LRS secondary eclipses and (right) residuals after subtracting the best-fit eclipse and instrumental ramp model. The strong wavelength-dependent instrumental ramp is visible at the beginning of the raw light curves for both observations.
\label{fig:waterfall}}
\end{figure*} 

Although photochemical hazes provide a plausible alternative to silicate clouds at lower temperatures, these hazes are unlikely to be present in significant quantities in HD~189733~b's dayside or terminator regions. \cite{gao_aerosol_2020} predict that photochemically produced hazes should become the dominant aerosol for planetary atmospheres cooler than 950 K. This is confirmed by lab experiments which suggest hazes should be commonplace in atmospheres $\lesssim$800 K \citep{horst_haze_2018,he_laboratory_2018}. At the higher temperatures typical of HD~189733~b's day side and terminator, these hazes should be destroyed faster than they can accumulate. This suggests that silicate clouds remain the most plausible candidate for the aerosols observed in HD~189733~b's atmosphere.

Mid-infrared observations with Spitzer have confirmed the presence of silicate clouds in the atmospheres of isolated brown dwarfs at similar temperatures to hot Jupiters \citep[e.g.,][]{cushing_spitzer_2006,burgasser_clouds_2008,burningham_cloud_2021,suarez_ultracool_2022}. With JWST, we now have the ability to detect silicate absorption features in the atmospheres of transiting planets. Indeed, silicate cloud features have been recently confirmed with JWST/MIRI LRS in the atmospheres of both WASP-17 b \citep{Grant2023} and WASP-107~b \citep{dyrek_so2_2023}. JWST observations with MIRI LRS provide a means to confirm the presence of aerosols in the atmosphere of HD~189733~b and determine their composition for the first time by directly detecting their mid-infrared absorption features.

In this paper, we present an emission spectrum of the benchmark hot Jupiter HD~189733~b derived from two secondary eclipse measurements with JWST's MIRI instrument. We first discuss our observations and data reduction process (Sec.~\ref{sec:data}). We then compare our dayside emission spectrum to models and find an additional absorption feature at 8.7~$\mu$m, as well as a region around 9.6~$\mu$m that is poorly matched by our models. We discuss the robustness of these features between two different data reduction pipelines, \texttt{Eureka!} and \texttt{SPARTA}, and two secondary eclipse observations (Sec.~\ref{sec:model_comp}). We present possible explanations for both wavelength regions (Sec.~\ref{sec:self_consistent_grid} and ~\ref{sec:9.6_region}), and discuss their implications for the atmosphere of HD~189733~b in context with other JWST MIRI LRS observations of transiting giant planets (Sec.~\ref{sec:discussion}).

\section{Observations}

We observe two full secondary eclipses of HD~189733~b with the JWST MIRI LRS instrument \citep{rieke_mid-infrared_2015} operating in slitless (SL) mode during separate visits as part of GO program 2021 (PI Brian Kilpatrick). The first visit began on October 18, 2022 16:38:42 UT and ended on Oct 18, 2022 22:29:04 UT. The second visit began on Jun 30, 2023 19:49:47 UT and ended on July 1, 2023 01:41:22 UT. We utilized the SLITLESSPRISM subarray for our observations and performed target acquisition using the FND filter. We observed with MIRI in FAST read mode with NGroups = 5, corresponding to a total integration time of 0.95 s per integration. Each secondary eclipse observation included a total of 17,035 integrations, corresponding to a total length of approximately six hours not including overheads. Due to the brightness of our target, the 26 shortest-wavelength pixels reach 80\% of full-well within 5 groups, corresponding to wavelengths shorter than 5.43~$\mu$m.

\begin{deluxetable*}{cccccc}
\tablecaption{Fitted and fixed parameters for our \texttt{Eureka!} white light curve ($5-12$~$\mu$m) fits and retrieved values for each secondary eclipse fit both separately and jointly. Uniform priors and their bounds are denoted $\mathcal{U}$(a,b), and Guassian priors are denoted $\mathcal{N}$($\mu$,$\sigma$), where $\mu$ is the peak of the distribution, and $\sigma$ is the width.
\label{table:fitted_params}}
\tabletypesize{\footnotesize}
\tablehead{
\colhead{Parameter} & \colhead{Fixed/Free} & \colhead{Prior} & \colhead{Eclipse 1} & \colhead{Eclipse 2} & \colhead{Joint}}
\startdata
\multicolumn{6}{c}{Astrophysical Model} \\
\hline
Orbital period, P$^a$ & Fixed & -- & 2.218575 &  2.218575 & 2.218575 \\
Orbital eccentricity, $e$ & Fixed & -- & 0 & 0 & 0 \\
Planetary radius, $R_p$/$R_*$ & Free & $\mathcal{N}$(0.155,0.0001)$^b$ & $0.14\pm0.02$ & $0.17\pm0.03$ & $0.16\pm0.02$ \\
Semi-major axis, $a$/$R_*$ & Free & $\mathcal{N}$(8.89,0.01)$^b$ & $8.4\pm0.8$ & $9.4\pm0.9$ & $9.1\pm0.5$ \\
Inclination, $i$ [degrees] & Free & $\mathcal{N}$(85.72,0.02)$^b$ & $85\pm1$ & $86\pm1$ & $85.9\pm0.6$ \\
Eclipse depth, $F_p$/$F_*$ & Free & $\mathcal{U}$(0.001,0.006) & $0.003156\pm0.000009$ & $0.003132\pm0.000009$ & $0.003143\pm0.000006$ \\
Eclipse mid-time, $T_c$ [BMJD\_TDB] &  Free & t$_0$ + $\mathcal{U}$(0.07,0.15) & $59870.85625\pm0.00004$ & $60125.99235\pm0.00004$ & \\
\hline
\multicolumn{6}{c}{Systematics Model} \\
\hline
y-offset, $c_1$ & Free & $\mathcal{U}$(-1,1) & $0.025\pm0.001$ & $0.028\pm0.001$ &\\ 
x-offset, $c_2$ & Free & $\mathcal{U}$(-1,1) & $0.001\pm0.001$ & $0.002\pm0.001$ & \\ 
PSF-width, $c_3$ & Free &$\mathcal{U}$(-1,1) & $-0.04\pm0.001$ & $-0.04\pm0.01$ &\\ 
Linear time coefficient, $c_4$ & Free & $\mathcal{U}$(-1,1) & $-0.0032\pm0.0001$ & $-0.0010\pm0.0001$ & \\ 
Exponential amplitude, $c_5$ & Free & $\mathcal{U}$(-0.01,0.01) & $0.00115\pm0.00004$ & $0.00042\pm0.00006$ & \\
Exponential decay constant, $c_6$ & Free & $\mathcal{U}$(0,0.2) & $0.014\pm0.001$ & $0.010\pm0.001$ & \\ 
error scale & Free & log-$\mathcal{U}$(-8,-9) & $-8.2\pm0.03$ & $-8.2\pm0.03$ & \\
\hline
\enddata
\tablecomments{Our retrieved systematic model and eclipse center times from our joint fit are consistent with our individual eclipse fits, so we only show the individual fits.\\
    a - from \cite{ivshina_tess_2022} \\
    b - The Gaussian priors used for these values come from the best-fit values to Spitzer 8~$\mu$m transits of HD~189733 b from Table 4 in \citet{agol_climate_2010}.}
\end{deluxetable*}

\section{Data Reduction} \label{sec:data}

Following the precedent set by other JWST time series observations, \citep[e.g.,][]{CO2_ERS,kempton2023,Grant2023}, we performed two independent reductions of our data using the \texttt{Eureka!} \citep{Bell2022} and \texttt{SPARTA} \citep{kempton2023} pipelines in order to compare how different reduction choices affect the final spectrum. In the sections below we describe our general data reduction process and highlight key distinctions between the two pipelines.

\begin{figure*}[t]
\epsscale{1.}
\plotone{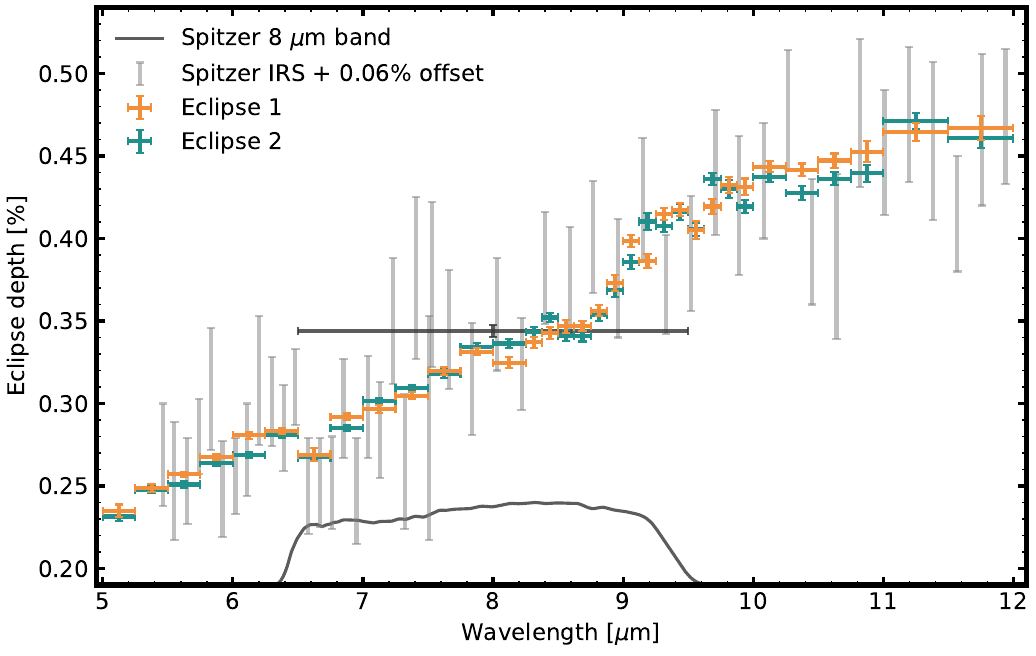}
\caption{Our fitted spectroscopic secondary eclipse depths from $5-12$~$\mu$m are plotted for both visits from our \texttt{Eureka!} reductions. The Spitzer 8 $\mu$m band response curve is shown in black. We compare our MIRI LRS emission spectrum to the Spitzer IRS spectrum from \cite{todorov_updated_2014} (grey), with an applied offset of 0.06\%. The misfit between eclipses at $10-11$~$\mu$m matches the area of the detector where the ramp behaviour differs the most significantly between visits (see Sec~\ref{sec:red_comp}). 
\label{fig:eclipse_depths}}
\end{figure*} 

\subsection{\texttt{Eureka!} reduction}

We began by using the open-source \texttt{Eureka!} \citep{Bell2022} pipeline to reduce the MIRI observations. We used version 1.8.2 of the \texttt{JWST} pipeline \citep{bushouse_2022}, \texttt{CRDS} (Calibration Data Reference System) version 11.16.20, and version 0.9 of \texttt{Eureka!}. We followed all the default steps of \texttt{Eureka!} as described in \cite{Bell2022}. By default, \texttt{Eureka!} utilizes the \texttt{JWST} pipeline Stage 1 steps of ramp fitting, non-linearity, gain scaling, saturation and data quality flagging. For the jump detection step, we increased the detection threshold to 16$\sigma$ to avoid introducing additional noise into our 1D spectra for this bright target, particularly at our shortest wavelengths where we had only 2-3 groups before saturation.

We also followed the default Stage 2 steps of the \texttt{Eureka!} pipeline, including the flat fielding and srctype steps of the JWST Science Calibration Pipeline. This meant that we skipped the background subtraction and flux calibration steps of the JWST Stage 2 pipeline in favour of the custom background subtraction and extraction of \texttt{Eureka!} Stage 3. This absolute flux calibration is unnecessary for our analysis, and has been shown to increase noise in the final light curves. We then passed the Stage 2 outputs through the \texttt{Eureka!} Stage 3 spectral extraction, which performs background subtraction and outputs 1D spectra from the 2D calibrated images. 

Due to the brightness of our target, we compared both optimal and standard box extraction. For both extraction methods, we masked pixels marked as “DO NOT USE” in the data quality array as well as any other unflagged pixels with ''NaN'' or ''inf'' values for the spectral extraction. We manually set the gain value to 3.1 as per \citet{Bell2022}. We performed background subtraction at each wavelength element by fitting a linear function in the cross-dispersion direction, considering only pixels outside an aperture with a half-width of 12 pixels centered on the trace. Similar to other MIRI LRS studies, we found this was sufficient to remove the observed periodic background noise \citep{bouwman_spectroscopic_2023,kempton2023}. For the spectral extraction step, we used an aperture with a half-width of 5 pixels centered on the fitted central pixel of the trace. We removed outliers using a double-iteration rejection scheme, flagging pixels deviating more than 5$\sigma$ from the median frame. We found that our choice of extraction method had a negligible effect on our final dayside emission spectrum, and therefore elected to use optimal extraction for our final analysis.

\subsection{\texttt{SPARTA} reduction}

To validate our \texttt{Eureka!} reduction, we also extracted spectra using the fully independent \texttt{SPARTA} pipeline, first referenced in \cite{kempton2023}. Unlike \texttt{Eureka!}, which uses the default \texttt{JWST} pipeline for Stage 1 and Stage 2 of the reduction, \texttt{SPARTA} uses a custom routine for fitting a slope to the ramp, described in \cite{kempton2023}, that accounts for the non-linear response of groups. This routine calculates custom weights for individual groups to straighten out non-linearity consistent across integrations resulting from sources such as the reset switch charge decay or last-frame effect. It additionally removes significant outliers that can result from cosmic rays. The non-linearity correction ignores the first group, leaving a total of 4 groups for this observation.

After the custom non-linearity correction in Stage 1, we used \texttt{SPARTA} to perform similar steps to those of the \texttt{JWST} pipeline Stage 2 reduction including flat fielding, gain correction, bad pixel mask and dark subtraction. For consistency we used the same reference files as for our \texttt{Eureka!} analysis to minimize potential sources of discrepancies between the two reductions. As in the \texttt{Eureka!} reduction, we use the constant gain value of 3.1 \citep{Bell2022}. Background subtraction is done by subtracting a row-wise median calculated using windows located above and below the trace, each with a width of 15 pixels and located 12 pixels from the center of the trace. We performed both optimal and standard extraction using a window with a half-width of 5 pixels centered on the trace. As with our \texttt{Eureka!} reduction, we compare both types of spectral extraction and find that they perform equivalently well, so we adopt the optimal extraction to compare with our \texttt{Eureka!} reduction.

\begin{figure*}[t!]
\epsscale{1.1}
\plotone{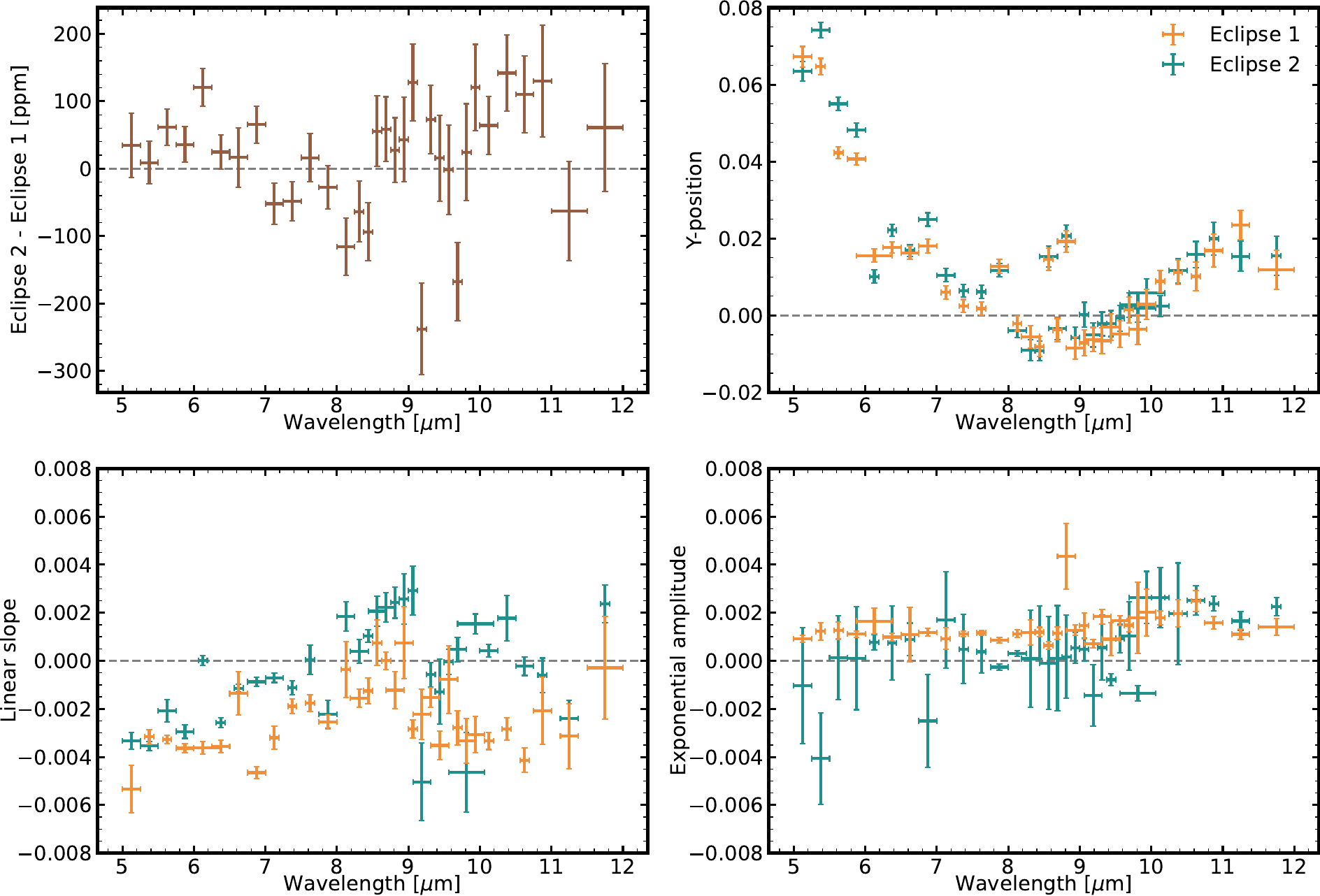}
\caption{The fitted spectroscopic coefficients for three of our detrending parameters, PSF width, linear slope with time, and ramp amplitude, for each of our two secondary eclipse observations from our \texttt{Eureka!} reduction.
While the best fit systematics differ between our two secondary eclipse observations, the shape of the emission spectrum is mostly consistent between the two visits at the $2\sigma$ level. The largest differences between visits occur where the linear component of the ramp varies most. The linear component of the ramp also switches sign between visits from $10-11$ $\mu$m. This suggests that this could lead to small offsets in eclipse depth (see Sec.~\ref{sec:red_comp}).}
\label{fig:sys_params}
\end{figure*} 

\subsection{Light Curve Fitting}

\begin{figure*}[t!]
\epsscale{1.}
\plotone{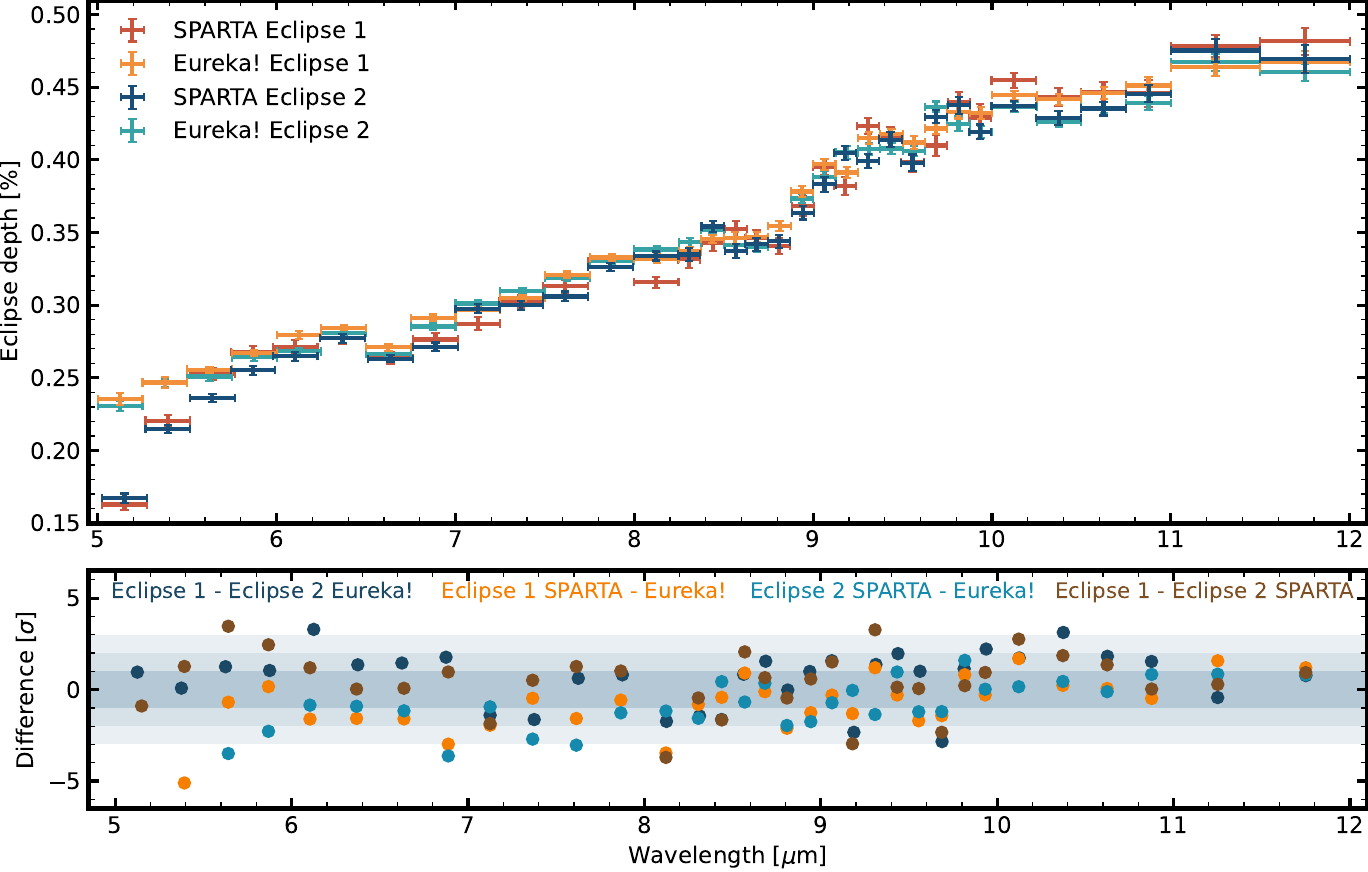}
\caption{Comparison of our \texttt{Eureka!} and \texttt{SPARTA} reductions for each of our two secondary eclipse observations. The systematic differences between reductions are due to the different linearity corrections. Shortwards of 5.5 $\mu$m, the difference is increased by the partially saturated detector. \label{fig:reduction_comp}}
\end{figure*} 

We took the extracted 1D spectra from both reduction pipelines and produced spectroscopic light curves by binning into 32 bins with wavelengths between 5 and 12~$\mu$m. We used a custom binning scheme with bin widths of either 0.125, 0.25, and 0.5~$\mu$m depending on wavelength. We used 0.5~$\mu$m bins longwards of 10~$\mu$m, where the noise substantially increases due to the partially illuminated detector. We used 0.25~$\mu$m bins elsewhere except from $8.4-10$~$\mu$m, where we used 0.125~$\mu$m bins to better resolve that region of the spectrum. We removed outliers in our photometric time series for each bin by masking points greater than 5$\sigma$ above the median value of a moving boxcar filter with a width of 10 points. For each frame we fit for an $x$- and $y$-offset by cross correlating with a median spectrum, and fitting for the peak position of the point spread function of the trace respectively. We included these values as detrending parameters in our light curve fits.

In order to fit our light curves we used the secondary eclipse model function implemented in the light curve fitting package \texttt{batman} \citep{kreidberg_batman_2015}. We assumed a circular orbit, and fit for a planet/star flux ratio, $F_p$/$F_*$ and secondary eclipse mid-time. For our spectroscopic light curve fits, we fixed the orbital separation, $a/R_*$, planetary radius $R_p/R_*$ and inclination to the values from the white light curve fit, shown in Table~\ref{table:fitted_params}.

As in other MIRI LRS time series observations \citep{kempton2023,Grant2023}, we find that our light curves exhibit a large, wavelength dependent ramp with a structure that changes from short to long wavelengths. In Fig.~\ref{fig:LC_correct} we plot a subset of representative spectroscopic light curves from this reduction with the best fit model overplotted in black. Similarly to these studies, we observed a turnover in this ramp at longer wavelengths in our second secondary eclipse observation where the ramp switches from negative to positive slope. This effect is thought to be dependent on the illumination history of the detector and it's idle-recovery behaviour \citep{bouwman_spectroscopic_2023}.

To detrend the ramp structure, we fit a combination of a linear slope and exponential ramp in time. We additionally discarded the first 350 integrations ($\sim$6 minutes) to remove the steepest part of the ramp at the start of each observation, which is poorly fit by a single exponential. We also detrended against the $x$-position (dispersion direction), $y$-position (cross-dispersion), and the width of the PSF ($\delta$y). We subtracted the median $y$ and $x$ position from our position arrays, as well as the median PSF width. Our final systematics model took the form:
\begin{equation}
    S = 1+c_1 y +  c_2 x + c_3 \delta y + c_4t + c_5 exp(-(t-t_0)/c_6).
\end{equation}
where each of the $c_i$s represent the coefficients for each of our detrending parameters. The prior and best fit values for each parameter are summarized in Table~\ref{table:fitted_params}. 

Our combined model for fitting our light curves is as follows:
\begin{equation}
    F_{rel} = M_{pl} (t) \times S(t)
\end{equation}
where M$_{pl}$ is the \texttt{batman} secondary eclipse model. 

In Fig.~\ref{fig:waterfall} we plot the raw spectroscopic light curves and residuals for our \texttt{Eureka!} reduction for both secondary eclipses. Our wavelength-dependent secondary eclipse depth spectrum for our \texttt{Eureka!} reduction is shown in Fig.~\ref{fig:eclipse_depths}. A subset of our wavelength-dependent detrending parameters are shown in Fig.~\ref{fig:sys_params}. In order to increase the visibility of absorption features, we convert our secondary eclipse depths to the corresponding brightness temperature using a BT-Settl model \citep{allard_bt-settl_2014} corresponding to the best fit stellar parameters of HD~189733. We adopt the parameters from \cite{rosenthal_california_2021}: $T_{eff}$ = 5012 K, log(g)= 4.5, and metallicity, [Fe/H] = 0.04. We adopt the value of $R_p/R_*$= 0.155 from the best-fit value of 7 Spitzer 8~$\mu$m transits published in \citet{agol_climate_2010}. The BT-Settl model is consistent with the shape of our flux-calibrated in-eclipse stellar spectrum at the level of $1-2$ percent difference, using the most recent flux calibration reference file available. The larger difference occurs in the saturated region of the detector. We therefore use a stellar model in order to ensure that our estimate of the stellar flux in the saturated regions does not bias our calculated brightness temperatures. The stellar spectra from both visits show $\lesssim0.3\%$ difference, so we use the same model for both visits. This is confirmed by ground-based monitoring of HD~189733~A, which confirms that stellar brightness variations are minimal during the time period these data was taken (Henry, G., private communication). Our secondary eclipse depths and corresponding brightness temperatures for our \texttt{Eureka!} reduction are shown in Table~\ref{table:spect_depths}.

\begin{figure*}[t!]
\epsscale{1.1}
\centering
\plotone{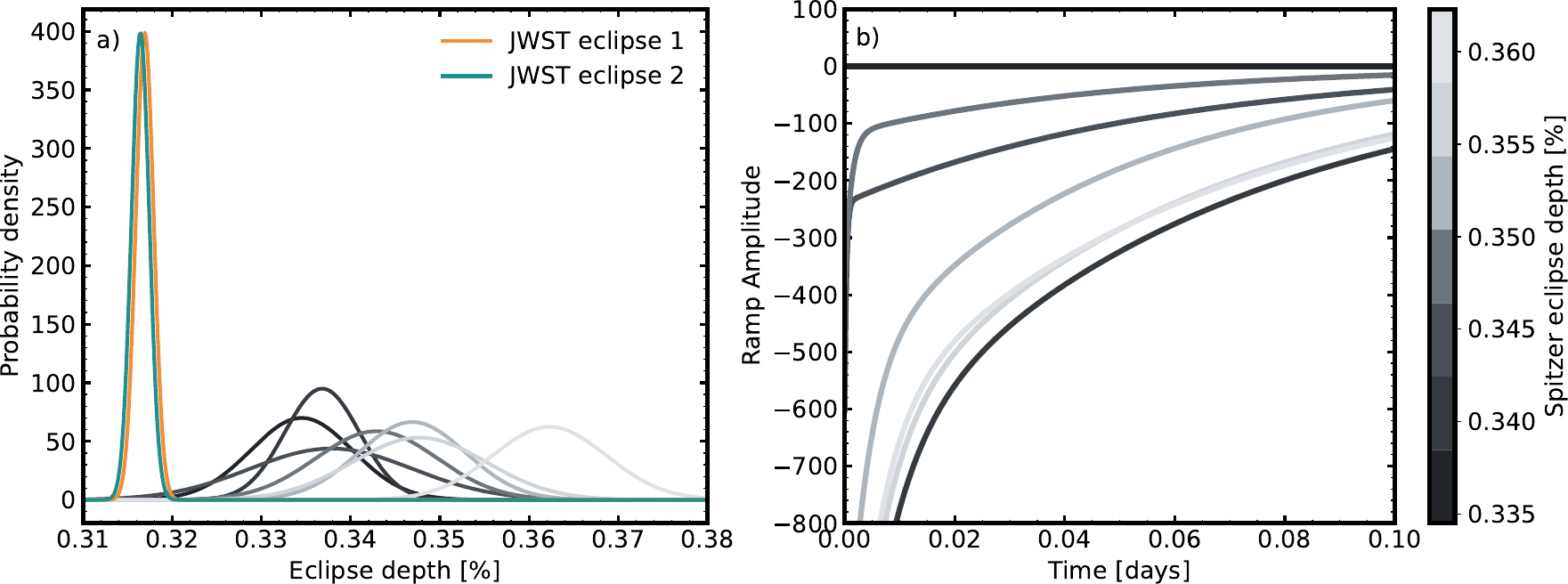}
\caption{a) Measured 8~$\mu$m eclipse depths colored by eclipse depth for the seven Spitzer eclipses from \cite{agol_climate_2010} as compared to our two JWST 8~$\mu$m eclipse depths (orange,blue). b) The ramps for the corresponding Spitzer eclipses from a) are plotted, colored according to their depths.
\label{fig:8mu}}
\end{figure*} 

\subsection{\texttt{Eureka!} and \texttt{SPARTA} intercomparison} \label{sec:red_comp}

In Fig.~\ref{fig:reduction_comp} we plot our spectroscopic secondary eclipse depths as a function of wavelength for each eclipse and data reduction pipeline. We find the largest deviation shortwards of 5.5 $
\mu$m. This occurs because most pixels are at least partially saturated in less than five groups short-wards of 5.4 $\mu$m, with the brightest pixels saturated in less than 3 groups. The small number of unsaturated groups results in a poor fit in the Stage 1 ``ramps-to-slopes" step, which will bias the final eclipse depth. This effect seems to be worse for the \texttt{SPARTA} reduction, which does not removed saturated groups. Reducing the number of groups used for extracting slopes in our \texttt{SPARTA} reduction to 3 does not improve the difference between reductions. We therefore used the \texttt{Eureka!} reduction for subsequent analyses. The difference in non-linearity treatment between the \texttt{SPARTA} and \texttt{Eureka!} pipelines results in the systematic offset between reductions. The shape of the spectrum is consistent between reductions beyond 5.5 $\mu$m within 3$\sigma$ therefore does not significantly impact any of our subsequent interpretation of spectroscopic features in HD~189733b's dayside spectrum.
 
We find that both of our secondary eclipse observations and reductions are consistent within 3$\sigma$. However, we do see a high number of spectral channels with differences at the 2$\sigma$ level than expected from just random noise. Since this also occurs for both eclipse observations between our two reductions, this suggests that our uncertainties are underestimated by a factor of $\sim$$2-3$ due to unknown systematics or other noise sources. One possible source is the additional correlation between the ramp shape and eclipse depth that we do not account for in our uncertainties. 

In Fig.~\ref{fig:sys_params}, we examine potential trends in the best-fit parameters from our systematics model as a function of wavelength. The shape of the time-dependent ramp, particularly the linear slope with time, is the most significant change in systematics between Eclipse 1 and Eclipse 2. The largest change occurs at the region between $10-11$~$\mu$m. We observe a transition in the ramp structure from an increasing linear slope in time shortwards of 10~$\mu$m to a downwards trend from $10-11$~$\mu$m in the light curves for Eclipse 2 that is not seen in Eclipse 1 (Fig.~\ref{fig:LC_correct}). Comparing the difference between our \texttt{Eureka!} reductions for Eclipse 1 and Eclipse 2 to the best fit ramp parameters for each visit reveals that the largest differences in secondary eclipse depth correspond well with the largest differences in the linear slope component of the ramp. This suggests that the divergence in the ramp amplitude and linear slope results in offsets in our retrieved secondary eclipse depth. This is further supported by observed covariances between the linear slope, exponential amplitude and eclipse depth within the posteriors for our spectroscopic light curve fits. The largest differences in linear slope appear to result in changes at the $\sim$$3\sigma$ level, which can be accounted for through inflation of our retrieved error bars. We see smaller differences in the ramp shape from $8.3-9$~$\mu$m between eclipses, where once again the ramp has a positive slope in Eclipse 2, and negative or zero slope in Eclipse 1. However that portion of the spectrum agrees well between visits, suggesting that only significant changes in ramp shapes as seen from $10-11$~$\mu$m bias the retrieved eclipse depth.

\begin{deluxetable*}{ccccc}
\tablecaption{Spectroscopic eclipse depths and brightness temperatures of HD~1897333 b from our \texttt{Eureka!} reduction for Eclipse 1 and Eclipse 2.
\label{table:spect_depths}}
\tabletypesize{\footnotesize}
\tablehead{
\colhead{Wavelength} & \colhead{Eclipse 1 depth} & \colhead{Eclipse 1 T$_{bright}$} & \colhead{Eclipse 2 depth} & \colhead{Eclipse 2 T$_{bright}$} \\ 
($\mu$m) & (ppm) & (K) & (ppm) & (K)}
\startdata
\hline
5.125 & $2356\pm40$ & $1231\pm4$ & $2305\pm35$ & $1221\pm4$\\
5.375 & $2467\pm36$ & $1236\pm4$ & $2464\pm27$ & $1235\pm3$\\
5.625 & $2556\pm21$ & $1229\pm2$& $2509\pm31$ & $1219\pm3$\\
5.875 & $2673\pm19$ & $1230\pm2$ & $2641\pm24$ & $1224\pm2$\\
6.125 & $2796\pm26$ & $1233\pm3$ & $2682\pm23$ & $1211\pm2$\\
6.375 & $2845\pm19$ & $1222\pm2$ & $2807\pm20$ & $1215\pm2$\\
6.625 & $2710\pm23$ & $1175\pm2$ & $2667\pm19$ & $1166\pm2$\\
6.875 & $2913\pm23$ & $1194\pm2$ & $2853\pm24$ & $1183\pm2$\\
7.125 & $2970\pm20$ & $1185\pm2$ & $3016\pm22$ & $1193\pm2$\\
7.375 & $3051\pm20$ & $1181\pm2$ & $3096\pm19$ & $1190\pm2$\\
7.625 & $3209\pm23$ & $1193\pm2$ &$3188\pm25$ & $1190\pm2$\\
7.875 & $3328\pm21$ & $1198\pm2$ & $3305\pm20$ & $1194\pm2$\\
8.125 & $3321\pm28$ & $1180\pm3$ & $3383\pm22$ & $1191\pm2$\\
8.3125 & $3370\pm31$ & $1177\pm3$ & $3432\pm30$ & $1189\pm3$\\
8.4375 & $3455\pm29$ & $1185\pm3$ & $3521\pm28$ & $1197\pm3$\\
8.5625 & $3461\pm41$ & $1179\pm4$ & $3414\pm38$ & $1171\pm3$\\
8.6875 & $3471\pm33$ & $1174\pm3$ & $3399\pm32$ & $1180\pm3$\\
8.8125 & $3544\pm36$ & $1179\pm3$ & $3546\pm33$ & $1161\pm3$\\
8.9375 & $3782\pm38$ & $1215\pm3$ & $3734\pm31$ & $1206\pm3$\\
9.0625 & $3970\pm38$ & $1242\pm3$ & $3880\pm42$ & $1226\pm4$\\
9.1875 & $3913\pm39$ & $1225\pm3$ & $4050\pm44$ & $1250\pm4$\\
9.3125 & $4150\pm40$ & $1260\pm3$ & $4075\pm38$ & $1245\pm3$\\
9.4375 & $4177\pm35$ & $1258\pm3$ & $4078\pm36$ & $1241\pm3$\\
9.5625 & $4119\pm44$ & $1242\pm4$ & $4062\pm36$ & $1232\pm3$\\
9.6875 & $4216\pm8$ & $1252\pm3$ & $4365\pm36$ & $1252\pm5$\\
9.8125 & $4328\pm43$ & $1266\pm4$ & $4249\pm54$ & $1278\pm3$\\
9.9375 & $4322\pm43$ & $1259\pm4$ & $4191\pm40$ & $1236\pm4$\\
10.125 & $4444\pm30$ & $1271\pm4$ & $4365\pm35$ & $1257\pm3$\\
10.375 & $4417\pm38$ & $1253\pm4$ & $4260\pm33$ & $1229\pm3$\\
10.625& $4461\pm42$ & $1251\pm5$ & $4360\pm37$ & $1235\pm3$\\
10.875 & $4512\pm60$ & $1256\pm3$ & $4392\pm50$ & $1231\pm4$\\
11.25 & $4636\pm56$ & $1258\pm5$ & $4672\pm58$ & $1234\pm5$\\
11.75 & $4676\pm73$ & $1246\pm6$ & $4604\pm59$ & $1264\pm5$\\
\enddata
\end{deluxetable*}

\subsection{Comparison with Spitzer Eclipses}

HD~1897333~b has been extensively studied in the mid-infrared with Spitzer. In Fig.~\ref{fig:eclipse_depths} we compare our JWST emission spectrum to the combined Spitzer IRS spectrum from \cite{todorov_updated_2014}. We allow for an offset, but other than that find good agreement between the slopes of the two spectra. The magnitude of the secondary eclipse depths in the Spitzer IRS data has been observed to shift depending on the correction used for the systematics and can vary significantly between individual reductions as a result \citep{grillmair_spitzer_2007,grillmair_strong_2008,todorov_updated_2014}. In order to be consistent with the broadband observations, \cite{zhang_platon_2020} found that a 30\% relative shift upwards was required in their retrievals. Therefore while we trust the slope of the overall spectrum, we consider the overall magnitude of the Spitzer IRS emission spectrum to be unreliable. 

The Spitzer IRS spectrum does not show a significant absorption feature at 8.7~$\mu$m, but this is expected given its relatively large measurement uncertainties. Like the Spitzer spectrum, our JWST spectrum shows an absorption feature from water at 6.5~$\mu$m. There is no obvious methane absorption in our spectrum, consistent with previous Spitzer observations. Both the IRS spectrum and our JWST spectrum show a dip from $10-11~\mu$m, though the magnitude of this feature varies with reduction method, and is smaller in our \texttt{Eureka!} Eclipse 1 spectrum. This is consistent with the location of absorption features from several silicate cloud features. It is also the location on the MIRI LRS detector where we observe changes in our systematics which may bias measurements of the eclipse depth (see the slopes of spectroscopic light curves shown in Fig.~\ref{fig:LC_correct}). 

In order to facilitate comparisons between our new measurements and the previous Spitzer 8~$\mu$m photometry, we bin our data to create eclipse light curves with the same wavelength coverage as the Spitzer 8~$\mu$m bandpass (See Fig~\ref{fig:eclipse_depths}). We account for the relative throughput between Spitzer and MIRI LRS and and fit for the corresponding secondary eclipse depths. In Fig.~\ref{fig:8mu}a we plot our 8~$\mu$m eclipse depths for both Eclipse 1 and 2 using our \texttt{Eureka!} reduction. Our two 8~$\mu$m JWST secondary eclipse depths are consistent within 1$\sigma$. We compare these two secondary eclipse depths to the seven 8~$\mu$m eclipse depths published in \cite{agol_climate_2010}. We find that our two 8~$\mu$m JWST secondary eclipse depths are both smaller than any of the seven eclipses from \citep{agol_climate_2010}, and differ from the average 8 $\mu$m eclipse depth reported in \cite{agol_climate_2010} by 280 ppm (3$\sigma$). HD 189733 A is an active star, with observed variations in brightness across a 17 year baseline of around 5-7\% based on ground-based photometric monitoring \citep[Henry, G., private communication;][]{sing_hubble_2011,knutson_36_2012}. Using Eq. 7 in \cite{zellem_forecasting_2017}, we find that this corresponds to a change in the secondary eclipse depth of as much as 220 ppm. We conclude that the offset between the Spitzer 8~$\mu$m secondary eclipse depths and our MIRI 8~$\mu$m secondary eclipse depths could potentially be explained by stellar variability.

We also note that the 8~$\mu$m photometry from Spitzer's IRAC instrument exhibited a ramp-like behavior similar to the one we see in our MIRI data. \cite{agol_climate_2010} found that their measured eclipse depths were sensitive to their choice of ramp model. They ultimately selected a double exponential model for their final eclipse depth measurements, as they found that this model resulted in the best agreement in eclipse depths across their sample of seven measurements. In Fig.~\ref{fig:8mu}b we plot the best-fit ramp model for each of the seven eclipses from \cite{agol_climate_2010}, colored by increasing depth. We find that all but one of their secondary eclipses observations show strong negative time dependent ramps. Within this ensemble, we see that increasing negative ramp strength correlates with a deeper measured eclipse depth. This suggests that the ramp shape may be systematically biasing the measured eclipse depths for all visits to values that are slightly higher than the true value in the sample from \cite{agol_climate_2010}. In our JWST data, we see a positive ramp at these wavelengths, which could in turn bias our measured eclipse depths to slightly lower values. The offset between these two ensembles of observations is therefore also possibly explained by instrumental systematics. This slight offset from either the ramp or stellar variability would manifest in a small shift in retrieved dayside temperature.

\section{Comparison to Models} \label{sec:model_comp}

We used two levels of model complexity and flexibility to interpret our measured mid-infrared dayside emission spectrum for HD~189733~b. First, we conducted a fully free chemistry, climate and cloud retrieval utilizing the retrieval code \texttt{petitRADTRANs}. The free retrieval framework allowed us to test different possible condensate species in order to limit the species included in our forward models to reduce computational time. Second, we used \texttt{picaso} to compute a grid of radiative-convective models that were self-consistent in climate and chemistry. This imposed the strongest physical constraint on the fitting process. We leveraged the forward model grid but additionally fit for aerosol parameters that were not physically self-consistent with the climate model grid. In this case, the cloud was able to form at any pressure and was not bound by the condensation curve of the condensate species. 

For both the free retrieval and grid we used a Bayesian framework to map our posterior space and determine our best fit parameters and errors. We fit both secondary eclipse measurements individually and jointly. The priors used for both fits are shown in Table~\ref{tab:grid}, as well as our median retrieved values and 1$\sigma$ uncertainties. For both our free retrievals and grid fits, we included an error inflation term, given the evidence that our uncertainties may be underestimated, as discussed in \ref{sec:red_comp}. We added a free error term added quadrature with the per point uncertainties obtained from our data reduction to obtain our combined error for our lileklihood function, $\sigma_i$ so our final per point error, $s_i$ took the form:
\begin{equation}
    s_i^2 = \sigma_i^2 + 10^b
\end{equation}
where b is the free error scale parameter.

\begin{deluxetable*}{cccccc}
\tablecaption{Retrieved parameters for HD~189733 b from free retrievals with \texttt{petitRADTRANS} and \texttt{Picaso} forward grid models on our \texttt{Eureka!} reduction. For our \texttt{petitRADTRANS} retrievals, molecules with firm upper limits are reported as 99\% upper limits.
\label{tab:grid}}
\tabletypesize{\footnotesize}
\tablehead{
\colhead{Parameter} &\colhead{Fixed/Free} & \colhead{Prior} & \colhead{Eclipse 1} & \colhead{Eclipse 2} & \colhead{Joint}} 
\startdata
\multicolumn{6}{c}{\texttt{petitRADTRANS} Free Retrievals} \\
\hline
$T_{eq}$ [K] & Free & $\mathcal{U}(1000,1400)$ & $1150\pm40$ &  $1170\pm60$ & $1110\pm50$  \\
$T_{int}$ [K] & Free & $\mathcal{U}(100,300)$ & $210\pm60$ & $210\pm60$ & $200\pm60$ \\
$\gamma$ & Free & $\mathcal{U}(0,1)$ & $0.38\pm0.05$ & $0.44\pm0.09$ & $0.33\pm0.06$ \\
log $\kappa_{IR}$ [K]  & Free & $\mathcal{U}(-6,2)$ & $-1.0\pm0.6$ & $-1.2\pm0.7$ & $-0.76\pm0.5$ \\
log $(g~\text{[cgs]})$ & Free & $\mathcal{U}(2.8,3.8)$ & $3.4\pm0.3$ & $3.4\pm0.3$ & $3.7\pm0.3$  \\
log $X_{H_2O}$ [MMR] & Free & $\mathcal{U}(-10,0)$ & $-3.0\pm0.8$ & $-3.0\pm1.0$ & $-3.7\pm0.6$  \\
log $X_{H_2S}$ [MMR] & Free & $\mathcal{U}(-10,0)$ & $-2.2\pm0.6$ & $<-0.9$ & $-2.4\pm0.5$  \\
log $X_{SO_2}$ [MMR] & Free & $\mathcal{U}(-10,0)$ & $<-4.2$ & $<4.3$ & $<-4.5$  \\
log $X_{CO_2}$ [MMR] & Free & $\mathcal{U}(-10,0)$ & $<-2.8$ & $<3.3 $ & $<-3.1$  \\
log $X_{SiO_2[s]}$ [MMR] & Free & $\mathcal{U}(-10,0)$ & $-3.6\pm0.6$ & $-3.9\pm0.8$ & $-3.6\pm0.7$\\
log $r_{SiO_2[s]}$ [cm] & Free & $\mathcal{U}(-8,-3)$ & $-6.5\pm0.7$ & $-6.6\pm0.8$ & $-6.5\pm0.7$\\
log $P_{SiO_2[s]}$ [bars] & Free & $\mathcal{U}(-6,2)$ & $-3.1\pm0.6$ & $-3.0\pm0.8$ & $-3.0\pm0.6$\\
$\sigma_{norm}$ & Free & $\mathcal{U}(1.0,3.0)$ & $1.9\pm0.6$ & $2.0\pm0.6$ & $1.9\pm0.5$\\
\hline
\multicolumn{6}{c}{\texttt{PICASO} Grid Fit W/O Clouds} \\
\hline
heat redis. &  Grid & [0.5,0.85] & $0.81^{+0.03}_{-0.06}$ & $0.79^{+0.04}_{-0.04}$ & $0.81^{+0.03}_{-0.04}$ \\
M/H [dex] &  Grid & [-1,1] & $-0.85^{+0.32}_{-0.1}$ & $-0.72^{+0.16}_{-0.16}$ & $-0.85^{+0.23}_{-0.1}$\\
C/O [rel. Sol=0.458] &  Grid & [0.25,2] & $0.38^{+0.27}_{-0.09}$ & $0.34^{+0.13}_{-0.06}$ & $0.36^{+0.15}_{-0.07}$\\
b [log K]] &  Free & $\mathcal{U}(-2.5,2.5)$ & $1.41^{+0.06}_{-0.05}$ & $1.42^{+0.06}_{-0.05}$ & $1.41^{+0.06}_{-0.05}$ \\
$\log$Z & - & - & -152 & -154 & -152 \\
\hline
\multicolumn{6}{c}{\texttt{PICASO} Grid Fit W/ Retrieved Clouds} \\
\hline
heat redis. &  Grid & [0.5,0.85] & $0.83^{+0.01}_{-0.02}$ & $0.82^{+0.02}_{-0.03}$ & $0.82^{+0.02}_{-0.02}$ \\
M/H [dex] &  Grid & [-1,1] & $-0.88^{+0.79}_{-0.09}$ & $-0.05^{+0.29}_{-0.19}$ & $-0.12^{+0.19}_{-0.16}$\\
C/O [rel. Sol=0.458] &  Grid & [0.25,2] & $0.74^{+0.77}_{-0.15}$ & $1.32^{+0.14}_{-0.44}$ & $1.43^{+0.06}_{-0.36}$ \\
f$_\mathrm{sed}$ [dex] &  Free & $\mathcal{U}(-1,1)$ & $-0.24^{+0.55}_{-0.51}$ & $-0.46^{+0.62}_{-0.37}$ & $-0.52^{+0.73}_{-0.37}$ \\
$\log \delta_{SiO_2[s]}$ [cm$^{-2}$] &  Free & $\mathcal{U}(1,10)$ & $4.43^{+1.27}_{-1.49}$ & $5.1^{+0.98}_{-1.16}$ & $5.33^{+0.79}_{-1.21}$ \\
$\log r_{SiO_2[s]}$ [cm] &  Free & $\mathcal{U}(-7,-3)$ & $-5.21^{+1.0}_{-1.26}$ & $-5.69^{+1.01}_{-0.91}$ & $-5.96^{+1.17}_{-0.75}$ \\
$\log P_{SiO_2[s]}$ [bar] &  Free & $\mathcal{U}(1,-4)$ & $-2.85^{+1.97}_{-0.87}$ & $-2.54^{+2.28}_{-1.1}$ & $-3.45^{+1.0}_{-0.41}$ \\
$\sigma_\mathrm{norm}$ - &  Free & $\mathcal{U}(0.5,2.5)$ & $0.77^{+0.17}_{-0.18}$ & $0.76^{+0.14}_{-0.17}$ & $0.8^{+0.11}_{-0.18}$ \\
b [log K]] &  Free & $\mathcal{U}(-2.5,2.5)$ & $1.13^{+0.06}_{-0.06}$ & $1.21^{+0.06}_{-0.06}$ & $1.16^{+0.07}_{-0.06}$ \\
$\log$Z & - & - & -128 & -135 & -131 \\
\hline
\enddata
\tablecomments{\texttt{PICASO} params: heat redis.= heat redistribution (full=0.5), M/H=log metallicity relative to solar, C/O=carbon-to-oxygen ratio relative to solar (solar=0.458), b=error inflation, f$\mathrm{sed}$=cloud sedimentation efficiency, $\log \delta_{SiO_2[s]}$=scaled cloud number density, $\log r_{SiO_2[s]}$=mean cloud particle radius, $\log P_{SiO_2[s]}$=cloud base pressure, $\sigma_\mathrm{norm}$=log normal radius distribution width for cloud particles, $\log$Z=log likelihood from fitting results)}
\end{deluxetable*}

\subsection{Free Bayesian Retrievals}

We perform free chemistry retrievals using the retrieval framework implemented in the open-sourced \texttt{petitRADTRANS} \citep{molliere_petitradtrans_2019}. Our retrievals included the free pressure temperature (P-T) profile parameterization from \cite{guillot_radiative_2010}, the surface gravity log($g$), and the planetary radius $R_P$ as free parameters, as well as gas phase species that might potentially be abundant enough to be detectable in the mid-infrared spectra of hot Jupiters, including H$_2$O, CO, CO$_2$, H$_2$S, FeH, SO$_2$, CH$_4$, C$_2$H$_2$, PH$_3$, HCN, NH$_3$ and SiO. We also include continuum opacity sources from H$_2$/H$_2$ and H$_2$/He collision-induced absorption. The sampling was performed using nested sampling implemented in \texttt{PyMultiNest}, the Python wrapper for MultiNest \citep{feroz_multinest_2009}, using 1000 live points. We evaluated the evidence with the sampling efficiency set to 0.3. We find an additional $95\pm10$ ppm error term added in quadrature is preferred in our free retrievals. This effectively increases our errors by a factor of $2-4$ depending on the channel.

We find a feature at 8.7 $\mu$m that we are unable to reproduce with the equilibrium gas phase absorbers predicted in the atmospheres of giant planets. We tested all of the gas-phase opacities available within \texttt{petitRADTRANS} and \texttt{ExoMol}, but were unable to fit these features. Next, we used Mie scattering coefficients, adapted from \cite{kitzmann_optical_2018}, to explore possible cloud species that might plausibly be present in the atmosphere of HD~1897333 b. We set up our cloud retrievals with the cloud base pressure, a power law scaling with pressure above the cloud base, cloud particle radius, and cloud abundance as free parameters. We tested scattering coefficients for MnS[s], ZnS[s], KCl[s], Fe[s], MgSiO$_3$[s], Mg$_2$SiO$_4$[s], and SiO$_2$[s] (where [s] refers to the solid phase). MgSiO$_3$[s] and Mg$_2$SiO$_4$[s] are thought to be the dominant silicate species in the atmospheres of hot Jupiters, but these species have strong features centered on 10~$\mu$m, not 8.7~$\mu$m \citep{wakeford_transmission_2015}. We considered both crystalline and amorphous particles when available in \texttt{petitRADTRANS}. We also fit a clear model and a simple grey cloud model with a cloud base and power-law opacity with pressure to compare with our cloudy models. 

\begin{figure*}[t!]
\epsscale{1.1}
\plotone{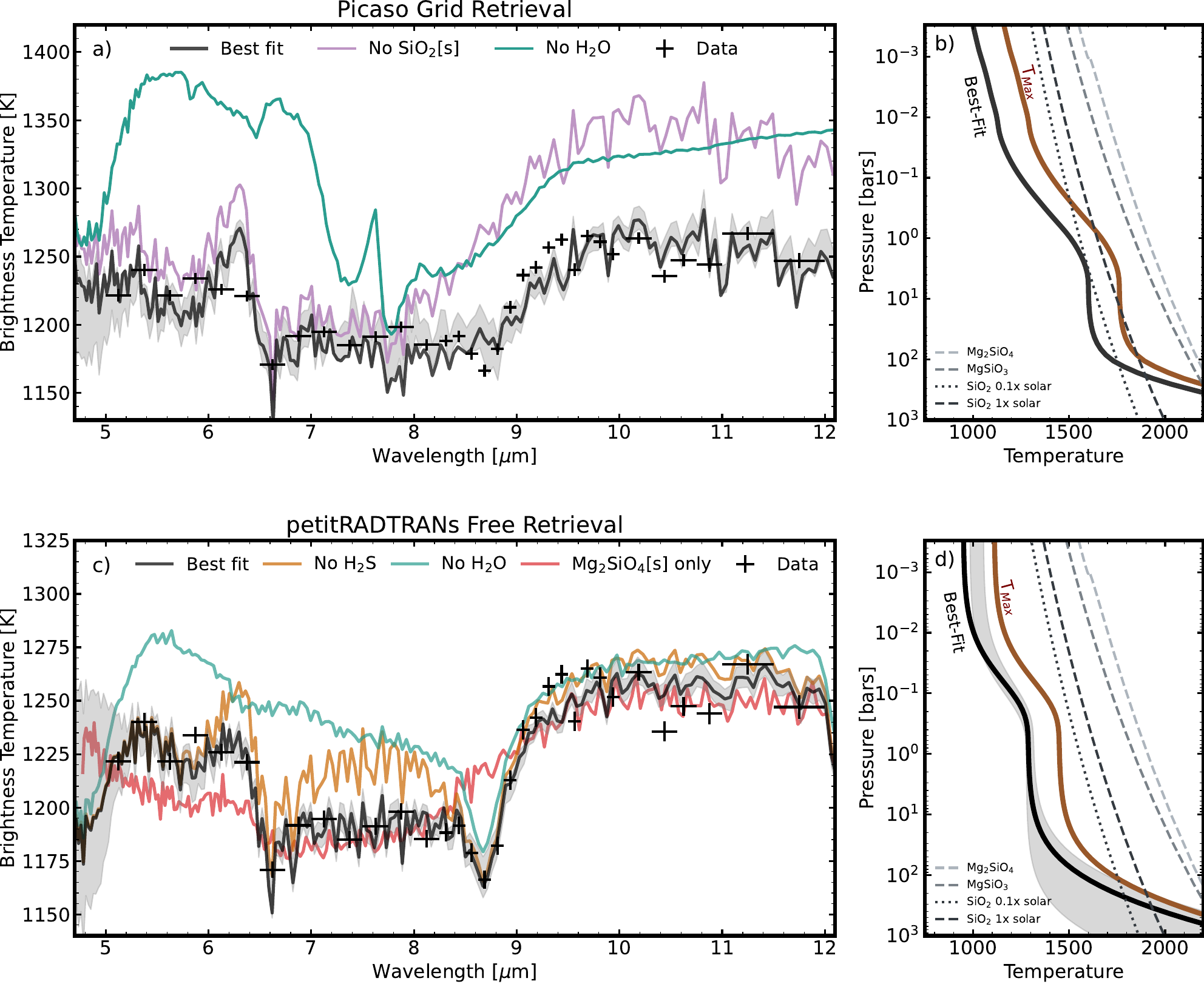}
\caption{a) Our joint eclipse spectrum, converted to brightness temperature, plotted against the best fit \texttt{Picaso} forward model and 3$\sigma$ upper limits. 
We show the same model with the contribution from the quartz (SiO$_2$[s]) grains (purple) and H$_2$O (blue) opacities removed. b) The best fit temperature-pressure profile predicted by forward models and maximum dayside temperature compared to condensation curves of major silicate cloud species. c) Our joint eclipse spectrum plotted against the best fit \texttt{petitRADNTRANs} free retrieval. We additionally compare to our best-fit model from a retrieval with fosterite (Mg$_2$SiO$_4$[s]) as the condensate species instead of SiO$_2$ (pink). We show the same model with opacities from the gas phase molecules detected in our retrievals, including H$_2$S (brown) and H$_2$O (blue), removed.
d) The best fit retrieved P-T profile from the free retrievals and 1-sigma limits compared to condensation curves.
\label{fig:T_bright}}
\end{figure*}

In Fig.~\ref{fig:T_bright}c, we plot our best-fit joint free retrieval against our secondary eclipse brightness temperature spectrum. Our best-fit P-T profile and 1$\sigma$ uncertainties are shown in Fig.~\ref{fig:T_bright}d.  We found that H$_2$O is the only gas phase molecule with a statistically significant detection in all of our fits, with a log-abundance of $-3.7\pm0.3$ (MMR), irrespective of the cloud model used in the fit. This value was consistent between eclipses, and is smaller than the measurement reported in \cite{finnerty_atmospheric_2023} using ground-based high-resolution spectroscopy with Keck/KPIC. We additionally detect H$_2$S in our first eclipse and our joint eclipse retrieval at a significance of 3.5$\sigma$. Our second eclipse retrieval only has a firm 99$\%$ upper limit of -0.9. H$_2$S has a broad, weak feature in the mid-infrared centered around 7 $\mu$m, which overlaps partially with the H$_2$O feature. As a result, it primarily affects the shape of the continuum and could be degenerate with clouds. However, the retrieved abundance of $-2.2\pm0.6$ (MMR) is consistent with equilibrium chemistry predictions based on the previously derived parameters for HD~1897333~b's atmospheric composition from \cite{finnerty_atmospheric_2023}. H$_2$S has also been detected in HD~189733~b's transmission spectrum \citep{fu_hydrogen_2024}, at an abundance of $-4.5^{+0.6}_{-0.4}$ (VMR), which is consistent at the 1.5$\sigma$ level with our retrieved value. Its presence in the dayside atmosphere of HD~189733~b can be confirmed by observations at shorter wavelengths with JWST. As in \citet{fu_hydrogen_2024}, we do not detect the photochemical product SO$_2$, retrieving only a strict upper limit of -4.7 (MMR). This is consistent with photochemical models for previously retrieved metallicity values for this atmosphere of $3-5\times$ solar \citep{fu_hydrogen_2024}.

We found that both amorphous and crystalline SiO$_2$[s] clouds were able to reproduce the absorption feature at 8.7~$\mu$m. Amorphous SiO$_2$[s] clouds were preferred only slightly to crystalline at $\lesssim1\sigma$. We used the Bayes factor to compare to both grey cloud and clear atmosphere models and found that SiO$_2$[s] clouds are preferred over the clear atmosphere model at 6.0$\sigma$, and over the grey cloud model at 6.1$\sigma$ for our joint eclipse retrievals. When compared to other condensate clouds, SiO$_2$[s] clouds are preferred to Mg$_2$SiO$_4$[s] at 6.1$\sigma$, MgiO$_3$[s] at 6.3$\sigma$, KCl[s] at 6.5$\sigma$ and ZnS[s] at 6.2$\sigma$. All condensate cloud species tested but SiO$_2$ produce a comparably good fit to the retrieval with Mg$_2$SiO$_4$[s] shown in Fig.~\ref{fig:T_bright}c as they fail to reproduce the 8.7 $\mu$m feature. We were unable to reproduce the spectral shape at 9.6~$\mu$m with any gas phase or scattering condensate opacity available in \texttt{petitRADTRANs} despite the flexibility of the free retrieval framework.

The largest difference between our \texttt{Eureka!} and \texttt{SPARTA} reductions occurs in the saturated wavelength range from $5-5.5$ $\mu$m. In order to make sure that this region is not biasing our fits, we perform additional retrievals excluding wavelengths shorter than 5.5 $\mu$m. We find no significant changes to our retrieved abundances or our retrieved cloud parameters. We also test the sensitivity of our retrieval results to our chosen binning scheme by re-binning our emission spectrum to uniform bin widths of 0.5, 0.25 and 0.125 $\mu$m. We then ran retrievals on each of these binning schemes and recovered consistent solutions. The only exception was our fit using 0.5 $\mu$m bins, where we found that the statistical significance of our water detection decreased due to our inability to fully resolve the water absorption feature at 6.5 microns.

\subsection{Self-Consistent Radiative Convective Model Grid}\label{sec:self_consistent_grid}

The radiative-convective thermochemical equilibrium (RCTE) grid was computed using \texttt{PICASO v3.0} \citep{Batalha_picaso2019}. The climate modeling functionality in \texttt{PICASO} is based on the legacy code developed to study Titan \citep{McKay1989_titan} and Uranus \citep{Marley1999_uranus}. The new open source code is fully described in \cite{mukherjee2023picaso}. Of note are the correlated-$k$ opacities used in \texttt{PICASO}, which are described in \citet{Marley2021} and available for download in \citet{lupu_roxana_2021_5590989}. The correlated-$k$ table is precomputed using a grid of chemical equilibrium abundances for different metallicity and C/O values at 1460 pressure-temperature points. The chemistry grid contains 29 species and is detailed in \citet{Marley2021} using NASA's CEA code \citep{gordon1994computer} and the modeling work detailed in \citet{fegleylodders1994}, \citet{lodders99}, \citet{lodders02}, \citet{LoddersFegley2002}, \citet{visscher06}, and \citet{visscher_atmospheric_2010}. Lastly, our chemical equilibrium grid relies on the elemental abundances of \citet{Lodders2010} where solar C/O=0.458.

The grid made for this analysis consisted of a total of 150 models of varying atmospheric properties at an interior temperature of 200 K and without radiatively active clouds. The properties varied were the atmospheric metallicity with six values from 0.1 to 10$\times$ solar, the C/O ratio with five values from 0.25 to 2$\times$ solar, and the heat redistribution factor with five values from 0.5 to 0.85. The heat redistribution factors specifically cluster around the higher values, with models generated at 0.5, 0.65, 0.75, 0.8, and 0.85 (a redistribution factor of 0.5 corresponds to full heat redistribution to the night side).

Our grid fitting procedure is based on the post-processing procedures utilized in studies of WASP-39~b \citep[e.g.,][]{Rustamkulov2023} and WASP-17~b \citep{Grant2023}, among others. In this framework, the pressure-temperature profile is physically constrained by the RCTE grid, as well as the chemistry, which is constrained by chemical-equilibrium abundances at each pressure layer. We leverage the open source sampling code \texttt{Ultranest} \citep{Ultranest}, which uses the MLFriends algorithm \citep{MLFriends2019} to fit for the best-fit parameters. This simply corresponds to our 3 grid parameters (M/H, C/O, and heat redistribution) for the cloud-free case. In doing so, we linearly interpolate the spectra computed for each grid point using \texttt{PICASO}'s rapid \texttt{custom\_interp} function. For our cloud-free grid fit, we obtain a single best-fit model that corresponds to M/H=1.5$\times$Solar, C/O=1.5$\times$Solar (=0.687 absolute), and heat redistribution=0.7 (see the full table of retrieved parameters in Table \ref{tab:grid}). 

In order to incorporate clouds we add five new free parameters to our fit: 1) the base pressure level of the cloud deck, 2) the sedimentation efficiency (f$_\mathrm{sed}$), 3) the cloud particle density, 4) the mean particle radius, and 5) the particle size distribution width assuming a log normal distribution. For our cloud fit we assume a cloud species of SiO$_2$[s] and use the $\alpha$-crystal optical constants computed at 928 K by \citet{zeidler2013optical} (T$_{eq}$ of HD~189733~b is $\sim$1191~K). Note that in the free retrieval section above, we test out several different cloud species. These five free parameters allow us to represent a cloud deck with particles that are log-normally distributed, similar to that of the parameterized cloud model of \cite{Ackerman2001}. In this model, the sedimentation efficiency, f$_\mathrm{sed}$, describes the vertical extent of the cloud deck. Low f$_\mathrm{sed}$ ($<$1) represents a cloud with a large vertical extent, and high f$_\mathrm{sed}$ ($>$1) represents a cloud that quickly becomes optically thin toward decreasing pressures. Our cloud routine is available via the \texttt{Virga} \citep{Virga,Rooney2022ApJ} open source tool function \texttt{calc\_optics\_user\_r\_dist} \footnote{\href{https://github.com/natashabatalha/virga/blob/dev/docs/notebooks/9_CustomCloudLayer.ipynb}{See \texttt{Virga} tutorial}}. Unlike using the \texttt{Virga} cloud model directly, this cloud fitting routine provides more physical flexibility by allowing the cloud to form away from the computed saturation vapor pressure curve for SiO$_2$ derived in \citet{Grant2023}. It also removes the physical relation between the particle radius, vertical mixing velocity (K$_\mathrm{zz}$), and f$_\mathrm{sed}$. 

The parameters derived from our grid fit are also available in Table \ref{tab:grid}. Overall the cloudy model is strongly preferred to the cloud-free model with a difference in the $\ln$ Bayes Factor$>$19 for both individual eclipse visits as well as the joint fit. This translates to preference for the cloudy models at $6-7\sigma$ \citep[as defined in Eqn. 21 from][]{Trotta2008}.
We find that the metallicity and C/O ratio are not well-constrained in our fits. This is not surprising, as we were only able to detect two gas phase molecules, H$_2$O and H$_2$S, in our free retrievals. As a result, there is a degeneracy between our retrieved M/H and C/O. We find that the H$_2$O volume mixing ratio (VMR) is well-constrained with a value of $\log$ H$_2$O = $-3.8^{+0.2}_{-0.1}$ for the cloudy model fits. This is consistent with the value found in our free retrievals, of $-4.1\pm0.4$ (VMR).

The inferred cloud parameters are consistent within 1$\sigma$ regardless of eclipse visit or joint fit. The cloud parameters are consistent with a vertically extended cloud deck starting at $\log P\sim-3$ [bar] with a mean particle radius of $<\log r=-4$ [cm]. For example, for the joint visit the retrieved cloud deck was -3.45$^{+1.0}_{-0.41}$ compared to -2.54$^{+2.28}_{-1.1}$, for eclipse 2. This level of agreement is also true for the cloud number density, particle radius, sedimentation efficiency, and particle distribution width. Therefore, the eclipse visit did not affect the inferred cloud parameters. 

We additionally considered whether the presence of SiO$_2$[s] grains in the atmosphere of HD~189733~b should affect the thermal structure of the atmosphere through radiative feedback. We calculated the total optical depth from the clouds as a function of wavelength and found that the optical depth is always less than 0.05 between $1-8$ $\mu$m. Therefore, the radiative feedback from these clouds should have a minimal effect on the planet's vertical temperature profile.

\subsection{Spectral Region Beyond 9.0~$\mu$m}\label{sec:9.6_region}

In Fig.~\ref{fig:T_bright}, we compare our best fit forward model and free retrieval models to our joint eclipse spectrum. We find that even after the inclusion of SiO$_2$ clouds, we still obtain a relatively poor fit to the measured spectral shape beyond 9~$\mu$m. In the \texttt{PICASO} grid retrieval, the best fit cloud model does not capture the steep gradient in brightness temperature from $8.6-9.5$~$\mu$m. In the \texttt{petitRADTRANS} free retrieval, the gradient is fit better (residuals $< 1\sigma$), until 9.4~$\mu$m where the model begins to deviate from the data. This occurs for retrievals on both the \texttt{Eureka!} and \texttt{Sparta} data reductions. These sequence of points differ at more than 2$\sigma$ from the best-fit models, even allowing for error inflation. Additionally, in both eclipses and both data reductions, we observe a potential feature at 9.6~$\mu$m, shown in Fig~\ref{fig:reduction_comp}. While this feature is narrow, it appears consistently in all data reductions and in both visits. It vanishes for bin width larger than 0.3~$\mu$m, where the feature is no longer resolved. The other region with multiple consecutive points differing by more than 2$\sigma$ from the best-fit models is between $10-11$ $\mu$m, where the ramp behavior changes on the detector and the differences between our two secondary eclipse observations are large.

The mismatch in the gradient from  $8.6-9.5$~$\mu$m would likely benefit from more complex cloud modeling (e.g. patchy clouds, non-log-normal particle distributions, or improved optical properties). Additionally, if the 9.6~$\mu$m feature is real, the included species in our models could also be affecting this region. We briefly considered what absorbers could affect these wavelengths. Using the high-throughput quantum chemistry framework presented in \citet{Zapata2023}, we explored possible molecular species that could affect the spectral shape in this region. Methanol (CH$_3$OH) is observed in many astrophysical contexts including solar system bodies and planet-hosting disks around other stars \citep[e.g.,][]{Drabek-Maunder2017, Booth2021}. Although methanol is expected to be important in shaping HD~189733b's overall chemistry, its steady-state abundance is predicted to be many orders of magnitude lower than species such as H$_2$O, CO$_2$, and CH$_4$ \citep{Moses2011}. The available high-temperature line lists for methanol currently focus on near-infrared wavelengths \citep[e.g.,][]{Alrefae2014} and do not allow for a robust determination of plausible CH$_3$OH abundances in HD~189733b's atmosphere from our JWST MIRI data. At this current time, no sufficient mid-infrared opacities for this molecule exist in any spectroscopy database, (e.g. HITEMP \citep{rothman_hitemp_2010}, ExoMol \citep{tennyson_exomol_2016}, suitable for modeling existing exoplanet spectra. Future work focused on the expansion of current high-temperature line list for methanol, such as those from \citet{ding_high-temperature_2019}, combined with additional observations of HD~189733~b's atmosphere at near-infrared wavelengths will be needed in order.

\section{Discussion} \label{sec:discussion}

In the previous section, we identified sub-micron grains of SiO$_2$[s] as the likely source of the excess absorption observed at 8.7~$\mu$m in HD~189733 b's dayside emission spectrum. Here, we examine the implications of this detection for the atmosphere of HD~189733~b, and place our detection in a broader context by comparing it to published detections of silicate grains with MIRI LRS for other hot Jupiters.

\subsection{Silicate Grains in the Atmosphere of HD~189733~b}

 While silicate absorption features have been directly observed in the emission spectra of brown dwarfs and planetary mass companions between $5-14$~$\mu$m using Spitzer IRS \citep[][]{burgasser_clouds_2008,burningham_cloud_2021,suarez_ultracool_2022}, they were only recently detected in the atmospheres of transiting planets using JWST \citep{Grant2023,dyrek_so2_2023}. As illustrated in Fig. \ref{fig:eclipse_depths}, MIRI LRS covers an equivalent $5-12$~$\mu$m spectral region with a significantly higher spectral precision than Spitzer IRS, making it an ideal instrument to search for spectroscopic features from silicate clouds. Our detection of absorption from SiO$_2$[s] grains in HD~189733~b's dayside emission spectrum provides the first direct confirmation of silicate particles in the atmosphere of HD~189733 b. This provides evidence that the scattering particles invoked in previous studies to explain this planet's optical scattering slope in transmission \citep[e.g.,][]{PONT_2008,wakeford_transmission_2015} may indeed be comprised of silicate materials, though the composition of particles in the terminator region may differ in composition from those detected on the hotter dayside.
 
In Fig.~\ref{fig:T_bright}, we plot the best-fit spatially-averaged dayside temperature profile inferred from fits to our MIRI emission spectrum and compare it to the condensation curves of various silicate species. This temperature profile predicts that equilibrium condensation of all of these species should occur at relatively high pressures, where their effect on the emission spectrum should be minimal. Our retrievals, on the other hand, require cloud particles present at $10^{-2}-10^{-3}$ bars. However, we know from both general circulation models \citep{showman_atmospheric_2009,kataria_atmospheric_2015} and previously published phase curve observations of HD~189733~b that the dayside temperatures vary by hundreds of K across the dayside hemisphere \citep{knutson_map_2007,knutson_36_2012,majeau_two-dimensional_2012,de_wit_towards_2012}. This means that silicates could reasonably condense at the pressures retrieved in our models in the hottest regions of the dayside atmosphere near the substellar point (see panel b in Fig.~\ref{fig:T_bright}). At our best fit cloud base pressures, we find that a reasonable temperature-pressure profile for the hottest regions of HD~189733~b's dayside atmosphere could cross the condensation curve for SiO$_2$ only if the abundance is moderately sub-solar. Condensation curves for abundances closer to solar place the cloud base deeper in the atmosphere around 0.1-1 bars. Sub-micron-sized particles also have relatively long settling times, and we would therefore expect vertical mixing to transport these small particles to even lower pressures than the predicted cloud base \citep{parmentier_3d_2013,lee_dynamic_2016}. Alternatively, the retrieved cloud parameters are also consistent with homogeneous or heterogeneous nucleation in the upper atmosphere, rather than equilibrium cloud condensation \citep[e.g.][]{helling_dust_2001,lee_dynamic_2016,powell_formation_2018}.

It is worth considering why we see SiO$_2$[s] in HD~189733~b's dayside atmosphere, and not any of the other commonly predicted silicate cloud species. Equilibrium chemistry condensation models predict the formation of enstatite (MgSiO$_3$) and fosterite (Mg$_2$SiO$_4$) clouds \citep{burrows_chemical_1999,visscher_atmospheric_2010}. Microphysical cloud formation models from \cite{gao_aerosol_2020} similarly predict that forsterite should be the dominant form of silicate clouds based on the relative nucleation potentials between the various silicate species. These predictions appear to be in good agreement with Spitzer observations of silicate absorption features in brown dwarf atmospheres, which are best-matched by models with enstatite and forsterite compositions \citep{cushing_spitzer_2006, luna_empirically_2021}. On the other hand, kinetic cloud models from \cite{helling_detectability_2006} predict that SiO2$_2$ grains should dominate in the atmospheres of brown dwarfs. This has since been supported by \cite{burningham_cloud_2021}, who find SiO$_2$[s] in addition to enstatite clouds in the atmosphere of an isolated brown dwarf. Dynamic cloud formation models for HD~189733~b from \cite{lee_dynamic_2016} predict that SiO$_2$ will condense on top of TiO$_2$ seed particles to form small grains ($<0.1$~$\mu$m), and that these small grains will dominate the aerosol distribution at low dayside latitudes and low pressures. In these models, \cite{lee_dynamic_2016} find that in the hottest regions of HD~189733~b's day side, the growth of other silicate particles is suppressed. However, their models also predict that MgSiO$_3$ and Mg$_2$SiO$_4$ should dominate in the terminator regions probed by transmission spectroscopy. This prediction is testable with the upcoming MIRI transmission spectroscopy of HD~189733~b (Program IDs GTO 1185, PI Thomas P. Greene, and GO 1633, PI Drake Deming), which will be sensitive to the composition of cloud species closer to the terminator.

\subsection{Comparison to other JWST MIRI Silicate Cloud Detections}

Two recent studies of exoplanet transmission spectroscopy with JWST MIRI LRS have also produced detections of silicate clouds in the atmospheres of WASP-17~b \citep{Grant2023} and WASP-107~b \citep{dyrek_so2_2023}. Emission spectroscopy with MIRI LRS and JWST NIRCam has also provided evidence that aerosols are potentially present in the dayside atmosphere of WASP-69~b \citep{schlawin_multiple_2024}, with silicates being one of the models explored to reproduce their data. In terms of atmospheric equilibrium temperatures, HD~189733~b (T$_{eq}\sim$1191~K) sits in between WASP-17~b (T$_{eq}\sim$1771~K) and WASP-107~b (T$_{eq}\sim$736~K) \citep{Southworth2011}, which provides an opportunity to explore the formation of silicate clouds across this important phase space. 

\citet{Grant2023} detected absorption from small particles (0.01~$\mu$m) of SiO$_2$ in their JWST MIRI LRS transmission spectrum of hot Jupiter WASP-17~b with a sharp spectral peak near 8.6~$\mu$m. The predicted peak dayside temperatures of HD~189733~b at the pressures probed in emission ($\sim$100~mbar) are very similar to the predicted temperatures along WASP-17~b's terminator at the pressures probed in transmission ($\sim$$1-10$~mbar) \citep[e.g.,][]{kataria_atmospheric_2016}. However, their cloud properties in these regions might still differ if they have different atmospheric compositions; \citet{Grant2023} highlight the dependence of the condensation curve for SiO$_2$ on atmospheric metallicity. WASP-17~b appears to have a super-solar atmospheric metallicity \citep[up to 100$\times$ solar;][]{Alderson2022, Grant2023}. Although we cannot place strong constraints on HD~189733~b's atmospheric metallicity from our JWST MIRI emission observations alone, previous studies of HD~189733~b's transmission and emission spectrum indicate that it is best-matched by an atmospheric metallicity that is 3-10$\times$ solar values \citep[e.g.,][]{zhang_platon_2020}. Potentially, HD~189733~b's lower atmospheric metallicity could cause SiO$_2$ to condense at lower temperatures as compared with WASP-17~b (see Figure~\ref{fig:T_bright}). 

\citet{dyrek_so2_2023} also recently detected evidence for sub-micron size silicate grains in their JWST MIRI LRS transmission spectrum of WASP-107~b. The $5-12$~$\mu$m transmission spectrum of WASP-107~b does not exhibit a strong $8.6-8.7$~$\mu$m feature explained best by SiO$_2$ alone. Instead \citet{dyrek_so2_2023} find that the broad absorption due to clouds in WASP-107~b's mid-infrared transmission spectrum is best explained by composite clouds that include amorphous MgSiO$_3$, SiO$_2$, SiO, with strong evidence for SiO[s] as the dominant silicate species. SiO[s] is likely thermally unstable on the substantially hotter dayside of HD~189733~b compared to WASP-107~b's terminator region \citep{lee_dynamic_2016}, which could account for their different cloud compositions. 

Although the exact properties of the silicate cloud species identified in the atmospheres of WASP-17~b, WASP-107~b and now HD~189733~b may differ, they all challenge extrasolar giant planet cloud formation and evolution theories. Thermochemical equilibrium cloud formation \citep[e.g.,][]{Marley2015} would predict that magnesium silicate clouds (MgSiO$_3$[s] or Mg$_2$SiO$_4$[s]) would be the primary cloud species to form in hot Jupiter atmospheres and that these clouds would subside to below the photosphere for cooler planets such as WASP-107~b. Non-equilibrium cloud formation models \citep[e.g.,][]{helling_detectability_2006} do predict that SiO$_2$[s] should form a substantial fraction of the material composition of clouds in extrasolar giant planets and brown dwarfs, but not necessarily at the atmospheric temperatures and pressures where we are now observing them. As highlighted in works such as \citet{lee_dynamic_2016,parmentier2016, kataria_atmospheric_2016,komacek2017,lines2018}, cloud formation and evolution in hot Jupiters is strongly shaped by large day-night temperature gradients that also give rise to strong vertical and horizontal mixing. This strong atmospheric mixing is likely able to sustain the population of sub-micron size silicate cloud particles observed in WASP-17~b, HD~189733~b, and WASP-107~b. Further observations of giant exoplanet atmospheres spanning a broad range of temperatures, compositions, and gravities with JWST MIRI will hopefully provide additional detections of specific cloud species and constraints on their average particle sizes that will prove valuable to further refine cloud formation and evolution theories and models.

\section{Conclusions}

We have presented an updated mid-infrared dayside emission spectrum of the hot Jupiter HD~189733~b spanning $5-12$~$\mu$m from two secondary eclipses observed with JWST/MIRI LRS. We find that our emission spectra from each of our two secondary eclipse observations are consistent within expectations of random noise from both our \texttt{Eureka!} and \texttt{SPARTA} data reduction pipelines. These results showcase the ability of JWST/MIRI LRS emission spectroscopy to identify new gas-phase absorption and cloud features in the atmospheres of transiting exoplanets at mid-infrared wavelengths. Our two secondary eclipse observations of HD~189733~b also demonstrate the repeatability of observations for relatively bright targets with JWST, while greatly improving upon previous Spitzer/IRS observations of this planet. We find that our new secondary eclipse observations are consistent with the shape of previous Spitzer/IRS mid-infrared spectroscopy of this planet when we apply a fixed offset to the IRS data. 
When we bin our spectra to match the 8~$\mu$m Spitzer bandpass, we find that our new eclipse depths are slightly smaller than the ensemble of seven published Spitzer 8~$\mu$m eclipse observations, likely also as a result of instrumental systematics related to the steep time-dependent ramps in the Spitzer data. Overall, we find good consistency between the Spitzer and JWST observations for this planet.

Our dayside emission spectrum confirms the previous detections of H$_2$O in the dayside of HD~189733~b by Spitzer, HST, and ground-based observations, and is consistent with expectations for a moderately super-solar metallicity hydrogen-dominated atmosphere. We find that the absorption feature observed at 8.7~$\mu$m is well-matched by models with high altitude, sub-micron, $1.09^{+0.14}_{-0.06}\times 10^{-3}$ $\mu$m SiO$_2$ grains, confirming previous indirect inferences of scattering aerosols in this planet's atmosphere. 
Even after accounting for the effect of clouds on the spectrum, we find that the measured spectral shape at wavelengths beyond 9~$\mu$m is not well-matched by our models. However, we were unable to identify any candidate cloud or gas phases absorbers that could improve the fit in this region using currently available line lists. CH$_3$OH could potentially produce absorption in this region, but does not have a complete line list at relevant temperatures and pressures. Future improvements in mid-infrared opacities and additional JWST secondary eclipse measurements from NIRCAM at shorter wavelengths should be able to confirm or disprove the presence of trace species such as CH$_3$OH.

These same observations can be used to produce 2D eclipse maps of HD~189733 b's dayside atmosphere; we will present these maps in a future study and compare them to previously published 8~$\mu$m Spitzer eclipse maps for this planet. These new spectroscopic eclipse maps will also allow us to probe variations in the dayside atmospheric temperature structure over a wide range of pressures, and potentially to determine the spatial extent of the SiO$_2$ clouds in HD~189733 b's dayside atmosphere. HD~189733~b will soon be observed with multiple JWST instrument modes in both transmission and emission (Programs GTO 1185, PI Greene, GO 1633, PI Deming). By combining these data sets with the mid-infrared observations presented here, we will be able to fully quantify the molecular inventory of the dayside and terminator regions of HD~189733 b's atmosphere. Transmission observations with MIRI LRS will allow us to see if the clouds on the dayside extend into the terminator, and if these clouds have the same composition as those on the day side, or if they transition to enstatite and forsterite as predicted by \cite{lee_dynamic_2016}. These observations can also be used to search for trace molecules produced by photochemistry \citep[e.g.,][]{Moses2011}, including their spatially varying abundances, and to quantify the effect of upward mixing on the observed methane abundance in HD~189733 b's upper atmosphere \citep[e.g.,][]{fortney_hot_2021}. This extensive suite of JWST observations will make HD~189733 b a key touchstone for studies of silicate cloud formation and disequilibrium chemistry in hot Jupiter atmospheres, building on its legacy as one of the best-studied hot Jupiters with Spitzer and HST. 

\section*{Acknowledgements}
This work is based on observations made with the NASA/ESA/CSA James Webb Space Telescope. The data were obtained from the Mikulski Archive for Space Telescopes at the Space Telescope Science Institute, which is operated by the Association of Universities for Research in Astronomy, Inc., under NASA contract NAS 5-03127 for JWST. These observations are associated with the program JWST-GO-2021.
Support for program JWST-GO-2021 was provided by NASA through a grant from the Space Telescope Science Institute, which is operated by the Association of Universities for Research in Astronomy, Inc., under NASA contract NAS 5-03127. Resources supporting this work were provided by the NASA High-End Computing (HEC) Program through the NASA Advanced Supercomputing (NAS) Division at Ames Research Center. N.E.B. acknowledges support from NASA’S Interdisciplinary Consortia for Astrobiology Research (NNH19ZDA001N-ICAR) under award number 19-ICAR19\_2-0041. Part of the research was carried out at the Jet Propulsion Laboratory, California Institute of Technology, under contract with the National Aeronautics and Space Administration. The JWST data used in this observation were obtained from the Mikulski Archive for Space Telescopes (MAST) at the Space Telescope Science Institute. The specific observations can be accessed from \href{http://dx.doi.org/10.17909/87p2-td95}{DOI 10.17909/87p2-td95}. Data products from this paper will be available in this Zenodo repository: \href{http://10.5281/zenodo.11238250}{10.5281/zenodo.11238250}.

\facilities{JWST(MIRI LRS), Spitzer(IRAC,IRS)}

\software{\texttt{Eureka!} \citep{Bell2022}, \texttt{SPARTA} \citep{kempton2023}, \texttt{Picaso} \citep{Batalha_picaso2019}, \texttt{Virga} \citep{Virga}, \texttt{petitRADTRANS} \citep{molliere_petitradtrans_2019},\texttt{numpy}, \citep{harris2020array}, \texttt{scipy} \citep{2020SciPy-NMeth}, \texttt{matplotlib} \citep{Hunter:2007}, \texttt{astropy} \citep{astropy:2013,astropy:2018,astropy:2022}, \texttt{emcee} \citep{foreman-mackey_emcee_2013}, \texttt{pymultinest}, \texttt{pycuba}, \citep{buchner_multi}, \citep{2020SciPy-NMeth}}


\bibliography{refs}{}
\bibliographystyle{aasjournal}

\end{document}